\documentclass[a4paper,fleqn,longmktitle]{cas-dc}
\usepackage[authoryear]{natbib}
\usepackage{lineno} 
\def\tsc#1{\csdef{#1}{\textsc{\lowercase{#1}}\xspace}}
\tsc{WGM}
\tsc{QE}
\tsc{EP}
\tsc{PMS}
\tsc{BEC}
\tsc{DE}
\def\ref@jnl#1{{\jnl@style#1}}

\def\aj{\ref@jnl{AJ}}                   % Astronomical Journal
\def\actaa{\ref@jnl{Acta Astron.}}      % Acta Astronomica
\def\araa{\ref@jnl{ARA\&A}}             % Annual Review of Astron and Astrophys
\def\apj{\ref@jnl{ApJ}}                 % Astrophysical Journal
\def\apjl{\ref@jnl{ApJ}}                % Astrophysical Journal, Letters
\def\apjs{\ref@jnl{ApJS}}               % Astrophysical Journal, Supplement
\def\ao{\ref@jnl{Appl.~Opt.}}           % Applied Optics
\def\apss{\ref@jnl{Ap\&SS}}             % Astrophysics and Space Science
\def\aap{\ref@jnl{A\&A}}                % Astronomy and Astrophysics
\def\aapr{\ref@jnl{A\&A~Rev.}}          % Astronomy and Astrophysics Reviews
\def\aaps{\ref@jnl{A\&AS}}              % Astronomy and Astrophysics, 
\def\azh{\ref@jnl{AZh}}                 % Astronomicheskii Zhurnal
\def\baas{\ref@jnl{BAAS}}               % Bulletin of the AAS
\def\bac{\ref@jnl{Bull. astr. Inst. Czechosl.}}
\def\caa{\ref@jnl{Chinese Astron. Astrophys.}}
\def\cjaa{\ref@jnl{Chinese J. Astron. Astrophys.}}
\def\icarus{\ref@jnl{Icarus}}           % Icarus
\def\jcap{\ref@jnl{J. Cosmology Astropart. Phys.}}
\def\jrasc{\ref@jnl{JRASC}}             % Journal of the RAS of Canada
\def\memras{\ref@jnl{MmRAS}}            % Memoirs of the RAS
\def\mnras{\ref@jnl{MNRAS}}             % Monthly Notices of the RAS
\def\na{\ref@jnl{New A}}                % New Astronomy
\def\nar{\ref@jnl{New A Rev.}}          % New Astronomy Review
\def\pra{\ref@jnl{Phys.~Rev.~A}}        % Physical Review A: General Physics
\def\prb{\ref@jnl{Phys.~Rev.~B}}        % Physical Review B: Solid State
\def\prc{\ref@jnl{Phys.~Rev.~C}}        % Physical Review C
\def\prd{\ref@jnl{Phys.~Rev.~D}}        % Physical Review D
\def\pre{\ref@jnl{Phys.~Rev.~E}}        % Physical Review E
\def\prl{\ref@jnl{Phys.~Rev.~Lett.}}    % Physical Review Letters
\def\pasa{\ref@jnl{PASA}}               % Publications of the Astron. Soc. of 
\def\pasp{\ref@jnl{PASP}}               % Publications of the ASP
\def\pasj{\ref@jnl{PASJ}}               % Publications of the ASJ
\def\rmxaa{\ref@jnl{Rev. Mexicana Astron. Astrofis.}}%
\def\qjras{\ref@jnl{QJRAS}}             % Quarterly Journal of the RAS
\def\skytel{\ref@jnl{S\&T}}             % Sky and Telescope
\def\solphys{\ref@jnl{Sol.~Phys.}}      % Solar Physics
\def\sovast{\ref@jnl{Soviet~Ast.}}      % Soviet Astronomy
\def\ssr{\ref@jnl{Space~Sci.~Rev.}}     % Space Science Reviews
\def\zap{\ref@jnl{ZAp}}                 % Zeitschrift fuer Astrophysik
\def\nat{\ref@jnl{Nature}}              % Nature
\def\iaucirc{\ref@jnl{IAU~Circ.}}       % IAU Cirulars
\def\aplett{\ref@jnl{Astrophys.~Lett.}} % Astrophysics Letters
\def\apspr{\ref@jnl{Astrophys.~Space~Phys.~Res.}}
\def\bain{\ref@jnl{Bull.~Astron.~Inst.~Netherlands}} 
\def\fcp{\ref@jnl{Fund.~Cosmic~Phys.}}  % Fundamental Cosmic Physics
\def\gca{\ref@jnl{Geochim.~Cosmochim.~Acta}}   % Geochimica Cosmochimica Acta
\def\grl{\ref@jnl{Geophys.~Res.~Lett.}} % Geophysics Research Letters
\def\jcp{\ref@jnl{J.~Chem.~Phys.}}      % Journal of Chemical Physics
\def\jgr{\ref@jnl{J.~Geophys.~Res.}}    % Journal of Geophysics Research
\def\jqsrt{\ref@jnl{J.~Quant.~Spec.~Radiat.~Transf.}}
\def\memsai{\ref@jnl{Mem.~Soc.~Astron.~Italiana}}
\def\nphysa{\ref@jnl{Nucl.~Phys.~A}}   % Nuclear Physics A
\def\physrep{\ref@jnl{Phys.~Rep.}}   % Physics Reports
\def\physscr{\ref@jnl{Phys.~Scr}}   % Physica Scripta
\def\planss{\ref@jnl{Planet.~Space~Sci.}}   % Planetary Space Science
\def\procspie{\ref@jnl{Proc.~SPIE}}   % Proceedings of the SPIE

\hypersetup{
	bookmarks=true,         % show bookmarks bar?
	unicode=false,          % non-Latin characters in Acrobat’s bookmarks
	pdftoolbar=true,        % show Acrobat’s toolbar?
	pdfmenubar=true,        % show Acrobat’s menu?
	pdffitwindow=false,     % window fit to page when opened
	pdfstartview={FitH},    % fits the width of the page to the window
	pdftitle={Forbidden oxygen emissions in the upper atmosphere of Mars},    
	pdfauthor={Raghuram},     % author
	pdfsubject={Subject},   % subject of the document
	pdfcreator={Latex},   % creator of the document
	pdfproducer={Texstudio}, % producer of the document
	pdfnewwindow=true,      % links in new PDF window
	colorlinks=true,       % false: boxed links; true: colored links
	linkcolor=blue,          % color of internal links (change box color with 
	citecolor=blue,        % color of links to bibliography
	filecolor=cyan,         % color of file links
	urlcolor=blue       % color of external links
}

\usepackage{setspace}

\begin{document}
\let\WriteBookmarks\relax
\def\floatpagepagefraction{1}
\def\textpagefraction{.001}
\shorttitle{Forbidden atomic oxygen emissions in the Martian dayside 
upper 
atmosphere}
\shortauthors{Raghuram, Jain, and Bhardwaj}
%\begin{frontmatter}

\title [mode = title]{Forbidden atomic oxygen emissions in the  Martian 
dayside upper atmosphere}                      
%\tnotemark[1,2]

%\tnotetext[1]{This document is the results of the research
%   project funded by the National Science Foundation.}

%\tnotetext[2]{The second title footnote which is a longer text matter
%   to fill through the whole text width and overflow into
%   another line in the footnotes area of the first page.}

\author[1]{Susarla Raghuram}[type=editor,
                        auid=000,bioid=1,
                        prefix=,
                        %role=Researcher,
                        orcid=0000-0002-1309-2499]
\cormark[1]
\ead{raghuramsusarla@gmail.com}
%\ead[url]{www.cvr.cc, cvr@sayahna.org}

\credit{Conceptualization, Methodology, Software and Interpretation}

\address[1]{Physical Research Laboratory,  Ahmedabad, 380009,  India.}
\address[2]{Laboratory for Atmospheric and Space Physics, University of 
Colorado Boulder, Boulder, CO, USA.}

\author[2]{Sonal Kumar Jain}[orcid=0000-0002-1722-9392]
\author[1]{Anil Bhardwaj}[orcid=0000-0003-1693-453X]
%\author[2,3]{CV Rajagopal}[%
%   role=Co-ordinator,
%   suffix=Jr,
%\ead{cvr3@sayahna.org}
%\ead[URL]{www.sayahna.org}

%\credit{Data curation, Writing - Original draft preparation}

%\address[2]{Sayahna Foundation, Jagathy, Trivandrum 695014, India}

%\author%
%[1,3]
%{Rishi T.}
%\cormark[2]
%\fnmark[1,3]
%\ead{rishi@stmdocs.in}
%\ead[URL]{www.stmdocs.in}

%\address[3]{STM Document Engineering Pvt Ltd., Mepukada,
%    Malayinkil, Trivandrum 695571, India}

\cortext[cor1]{Corresponding author}
%\cortext[cor2]{Principal corresponding author}
%\fntext[fn1]{This is the first author footnote. but is common to third
%  author as well.}
%\fntext[fn2]{Another author footnote, this is a very long footnote and
%  it should be a really long footnote. But this footnote is not yet
%  sufficiently long enough to make two lines of footnote text.}

%\nonumnote{This note has no numbers. In this work we demonstrate $a_b$
%  the formation Y\_1 of a new type of polariton on the interface
%  between a cuprous oxide slab and a polystyrene micro-sphere placed
%  on the slab.
%  }

\begin{abstract}
Recently, Nadir and Occultation for Mars Discovery (NOMAD) ultraviolet and 
visible spectrometer instrument on board the European Space Agency’s 
ExoMars Trace Gas Orbiter (TGO) simultaneously  measured {the} limb 
emission 
intensities  for both [OI] 2972 
and  5577 \AA\ (green) emissions in the dayside of 
Martian upper atmosphere. But the atomic oxygen red-doublet emission lines 
([OI] 6300 and 6364 \AA), which are expected to be observed along with [OI] 
5577 and 2972 \AA\ emissions, are  found to be absent in the NOMAD-TGO 
dayside observed  spectra. We aim to explore the photochemistry of all these 
forbidden atomic oxygen emissions ([OI] 2972, 5577, 6300, 6464 \AA) in the 
Martian daylight upper atmosphere and suitable {conditions} for the 
simultaneous detection of these emissions lines  in the dayside 
{visible} 
spectra. A photochemical 
model is 
developed to study the production and loss processes of O($^1$S) and 
O($^1$D), which are the respective excited states of green and red-doublet 
emissions, by incorporating various chemical reactions of different 
O-bearing species in the upper atmosphere of Mars. By reducing \cite{Fox04} 
modelled {neutral} density profiles by a factor of {2}, the 
calculated limb intensity profiles for  [OI] 5577 and 
2972 \AA\ emissions are found to be  consistent with the NOMAD-TGO 
observations. In this case, at altitudes 
below 120 km, our  modelled limb intensity for 
[OI] 6300 \AA\ emission is smaller by a factor 2 to 5 compared to that of  
NOMAD-TGO observation for [OI] 2972 \AA\ emission, and above this distance it 
is comparable with the upper limit of the 
observation. We studied various parameters which can influence the limb 
intensities of these atomic oxygen forbidden emission lines. Our calculated 
limb intensity for [OI] 6300 \AA\ emission,  when 
the Mars is at near perihelion and for solar maximum condition, suggests 
that all these forbidden emissions should be observable in the NOMAD-TGO  
{visible} spectra taken on the 
dayside of Martian upper atmosphere.  More 
simultaneous observations of forbidden atomic oxygen emission lines will 
help to understand the photochemical processes of oxygen-bearing species in 
the dayside Martian upper atmosphere.  
\end{abstract}

%\includegraphics{figs/grabs.pdf}

%    \item A photochemical model is developed to study forbidden atomic oxygen 
%	emissions in the Martian upper atmosphere
%
%    \item We studied the role of various parameters which influence 
%    the limb intensities of these emissions.
% 
%
%	\item All these forbidden emission lines should be observable 
%	in NOMAD-TGO dayside spectra, when Mars is at near perihelion.


\maketitle

\section{Introduction}
%\linenumbers
\onehalfspacing
%\doublespacing

Forbidden atomic oxygen emissions are the prominent features in the 
{visible} spectra of terrestrial atmospheres. These emissions   have 
been studied extensively from various ground- and space-based observatories. 
Observation of these emissions provides valuable information about the 
energy deposition and chemical processes in the upper atmospheres of 
terrestrial planets. About 5\% of excited atomic oxygen atoms produced in 
$^1$S state decay directly to the ground state ($^3$P) and produce [OI] 
2972 \AA\ emission, and the rest decay via $^1$D state that results in 
{the} emission at wavelength 5577 \AA\ (green line). Radiative decay of 
O($^1$D) to the ground state produces the red-doublet emissions at wavelengths  
6300 and 6364 \AA, provided it is not collisionally quenched by other species. 
Since both [OI] 2972 and 5577 \AA\ emissions are produced due to de-excitation 
of same electronic excited state ($^1$S),  detection of one of  these lines 
confirms the presence of other. Similarly, the observation of  
green line also indicates the presence of red-doublet emissions but the 
opposite {is not always true in case the O($^1$S) state is not produced 
in the atmosphere or totally 
deactivated by collisions.}

Several theoretical works were carried out to study the photochemistry of 
metastable atomic oxygen atoms in the Martian upper atmosphere. About 40 
years ago, \cite{Fox79} predicted the forbidden emissions of atomic oxygen 
in the upper atmosphere of Mars. \cite{Simon09} developed a kinetic model 
to study the seasonal variation of the dayside [OI] 2972 \AA\ 
emission intensity along with  other important emissions, such as 
CO$_2^+$ UV-doublet and CO Cameron bands. These modelling studies show that 
photodissociative excitation of CO$_2$ is the major source of O($^1$S) in 
the upper atmosphere of Mars. But the study of these emissions 
by \cite{Huestis10} shows that dissociative recombination of O$_2^+$ also 
plays an important role in producing O($^1$S) and  determining the 
[OI] 2972 and 5577 \AA\ emission intensities. Hence, it is suggested that the 
observed [OI] 2972 \AA\ emission intensity should be used to monitor the 
Martian ionosphere and not the ambient neutral temperature.  The calculations 
made by \cite{Gronoff12b} and \cite{Gronoff12a} showed that the uncertainties 
associated with modelling parameters can influence the calculated limb 
intensities of spectroscopic emissions. To understand the photochemistry of 
these forbidden emissions, \cite{Jain13b} accounted for several production and 
loss mechanisms  of O($^1$S) and O($^1$D) in the Martian upper atmosphere and 
modelled the atomic oxygen emission intensities for both solar maximum and 
minimum conditions. Recently, \cite{Gkouvelis18} have reported observations of  
[OI] 2972 \AA\ emission in the Imaging Ultraviolet Spectrograph (IUVS) onboard 
Mars Atmosphere and Volatile EvolutioN (MAVEN) satellite observed 
ultraviolet spectra and constrained the quantum yield for photodissociation 
of CO$_2$ producing O($^1$S) at Ly-$\alpha$ wavelength (1216 \AA) as about 8\%.

The observation of [OI] 2972 \AA\ emission has been done in the Martian upper 
atmosphere by several {spacecraft-borne} ultraviolet spectrometers 
starting 
from Mariner 6 to the recent MAVEN mission \citep{Stewart72a,Leblanc06,Jain15}.
But the  observation of [OI] 5577 \AA\ emission  {had never been} 
reported in 
the dayside of Martian upper atmosphere until the recent detection from Nadir 
and Occultation for Mars Discovery ultraviolet and visible spectrometer 
instrument on board the European Space Agency’s ExoMars Trace Gas Orbiter 
(NOMAD-TGO) by  \cite{Gerard20}. {Owing to its wide detection range 
(2000--6500 \AA), this spectrometer is capable of observing the 
four forbidden atomic oxygen 
emissions  ([OI] 2972, 5577, 6300, and 6364 \AA)  simultaneously 
in the Martian upper atmosphere.} NOMAD-TGO could observe  both [OI] 2972 and 
5577 \AA\ emissions simultaneously and the observed intensity ratio is used to 
determine the corresponding transition probabilities ratio of O($^1$S). 
However,  the emission features at [OI] 
red-doublet wavelengths were found to be absent in the NOMAD-TGO observed 
dayside
spectra \citep[at the 1$\sigma$ 
level,][]{Gerard20}.  We aim to study the photochemistry of these forbidden 
emission lines in the dayside of Martian upper atmospheres and explore 
{suitable 
conditions} to observe all these emissions simultaneously in the NOMAD-TGO 
spectra.  By accounting for  the important production and loss 
mechanism of O($^1$S) and O($^1$D) {in our  well utilized} 
photochemical model, we studied the emission processes of forbidden atomic 
oxygen emission 
lines 
in the dayside Martian upper atmosphere. The model inputs and calculations are 
presented in Section~\ref{sec:ips_cals}. We present the results of our model 
calculations and discussion in Section~\ref{sec:res_dis}. In 
Section~\ref{sec:sum_con}, this work is summarized and conclusions are 
drawn.

%\cite{Slanger06}
%--------------------------------------------------------------------
\section{Model inputs and calculations}
\label{sec:ips_cals}
\begin{figure}
	\centering
	\includegraphics[width=\linewidth]{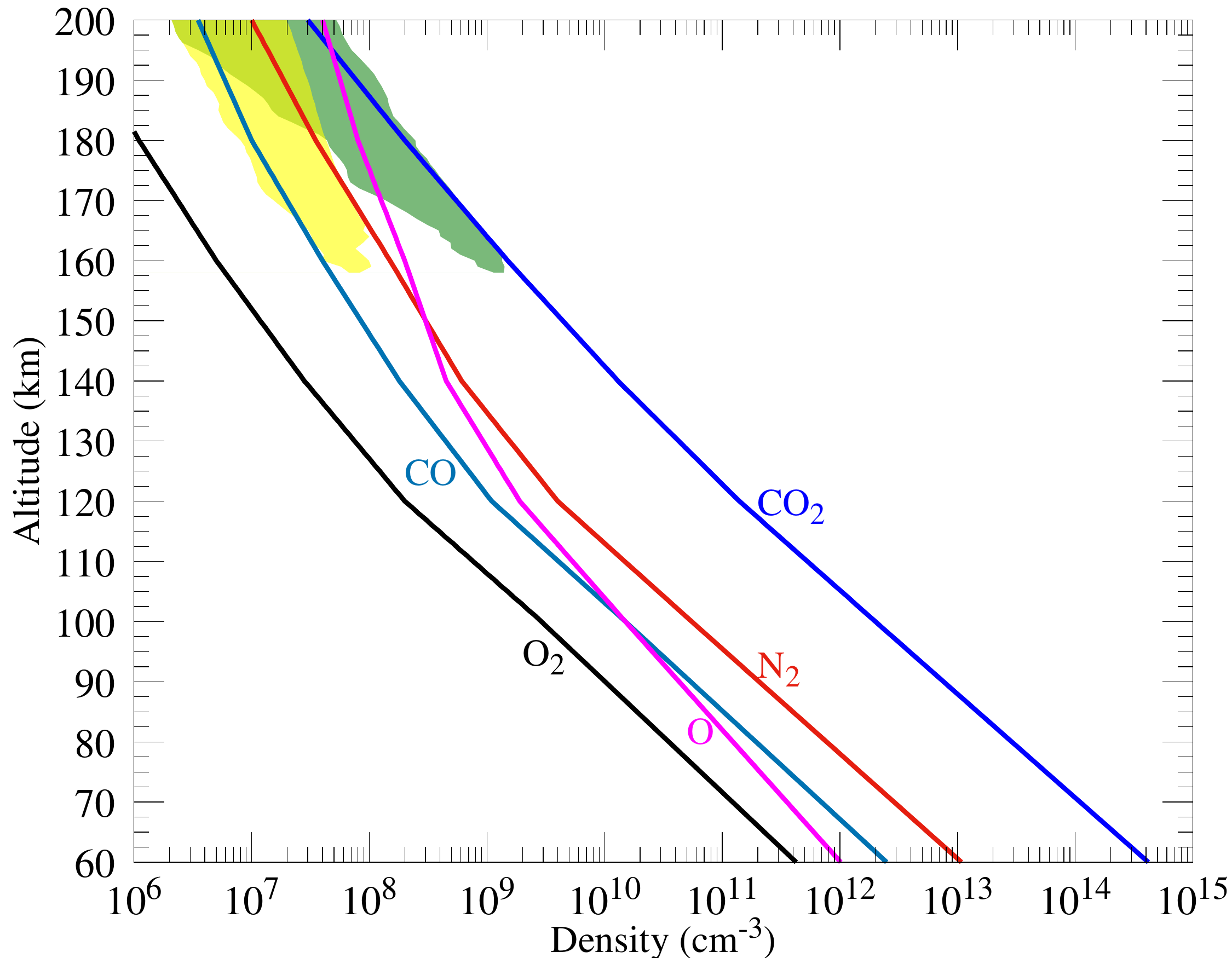}
	\caption{Neutral density distribution in the Martian upper atmosphere.
		The density profiles of major species are taken from \cite{Fox04}
		for solar minimum condition. 
		Green and yellow shaded areas represent the variability in the NGIMS 
		measured CO$_2$ and O densities during September 2019 for solar zenith 
		angle less than 40$^\circ$.}
	\label{fig:neudens}
\end{figure}

A detailed description of the model calculations  is provided in our earlier 
work \citep{Jain13b, Jain12, Raghuram12, Bhardwaj12, Raghuram20a}. Here we 
briefly describe the model inputs that are used for the present calculations. 
The neutral density profiles for the primary species (CO$_2$, CO, N$_2$, 
O$_2$, and O) of Martian upper atmosphere are considered from \cite{Fox04}, 
{which are based on the Viking measurements and for solar minimum 
condition.} To compare these  density profiles with the ongoing 
Neutral Gas and Ion Mass Spectrometer (NGIMS) on-board MAVEN mission  
measurements, we analysed NGIMS/MAVEN level 2 (L2), version 8, revision  1 
data for the period April to December 2019, during which NOMAD-TGO measured 
the limb emission intensities of forbidden oxygen emission lines in the Martian 
upper atmosphere. More details of the L2 data product are available in 
\cite{Benna18} and the data can be accessed from a web link 
(\href{https://pds-atmospheres.nmsu.edu}{https://pds-atmospheres.nmsu.edu}).
We noticed that most of the NGIMS/MAVEN measurements from April to December 
2019 were  on the night side of the Mars. Due to the MAVEN being on the night 
side, we do not have NGIMS/MAVEN measured neutral atmospheric densities at the 
time of NOMAD-TGO dayside observations. During September 2019,  observations 
were on the dayside Martian upper atmosphere for solar zenith angle (SZA) 
$<$40$^\circ$ and similar to the NOMAD-TGO observational conditions on 28 April 
2019 (SZA = 30$^\circ$) but at different Martian solar longitudes (L$_s$).  We 
plotted neutral density profiles of \cite{Fox04} along with variability in the  
NGIMS/MAVEN measured CO$_2$ and O density profiles in 
Figure~\ref{fig:neudens}. {The input solar radiation flux, which is a 
daily-averaged flux generated based upon on the Flare Irradiance Spectra 
Model-Mars (FISM-M) using the EUV calibrated band irrandiance measured by 
Solar Extreme Ultraviolet Monitor (EUVM) instrument onboard 
MAVEN and interpolated Earth-based solar 
indices, in the wavelength region of 5--1900 \AA\ is taken on {28 April 
2019} \citep[][\href{
https://pds-ppi.igpp.ucla.edu/search/view/?f=yes&id=pds://PPI/maven.euv.modelled
}{https://pds-ppi.igpp.ucla.edu}]{Eparvier15,Thiemann17}.}
As a case study,  we also calculated the limb intensity profiles for forbidden 
atomic oxygen  emissions for solar maximum condition by considering the neutral 
atmospheric model from \cite{Fox04}. For this calculation, we used EUVM onboard 
MAVEN measured solar flux on 12 December 2014 during which Mars was at 
perihelion.

The input photon and electron impact cross sections  are  described in 
\cite{Jain13b}. {Here we considered the total absorption cross section 
of CO$_2$ and the yield for the photodissociative 
excitation of CO$_2$ producing O($^1$S) from
\cite{Gkouvelis18}.} Based on the discussion of \cite{Gkouvelis18}, we have 
taken a quantum yield of 0.09 for the dissociative recombination of O$_2^+$ 
producing O($^1$S).
Following the assumptions of \cite{Gkouvelis18} and \cite{Jain13b}, we 
considered the quantum yields for the thermal recombination of CO$_2^+$ 
producing O($^1$S) and O($^1$D) as 0.05 and 0.59, respectively. However, we 
vary these assumed yields to study the impact of dissociative recombination of 
CO$_2^+$ on the modelled limb intensities. The photochemical reaction network 
used for  calculating various production and loss rates of O($^1$S) and O($^1$D) 
is presented in Table~\ref{Tab:reactions}. Electron and neutrals temperature 
profiles are taken from \cite{Fox09} for solar minimum and maximum conditions.

\renewcommand{\thefootnote}{\fnsymbol{footnote}}
\begin{table*}
	\centering
	\caption{Photochemical reactions for the formation and destruction of 
		O($^1$D) and O($^1$S) in the Martian upper atmosphere}
	\label{Tab:reactions}
	\begin{tabular}{llllllll}
		\hline
		No. & \multicolumn{2}{c}{Reaction} & Rate Coefficient & Reference \\
		\hline
		R1  & h$\nu$ + CO$_2$ & $\rightarrow$ O($^1$S) + CO &
		1.63 $\times$ 10$^{-7}$ & This  work \\
		R2  & h$\nu$ + CO & $\rightarrow$ O($^1$S) + C & 
		1.43 $\times$ 10$^{-8}$ & This  work \\
		R3  & h$\nu$ + O$_2$ & $\rightarrow$ O($^1$S) + O & 
		1.28 $\times$ 10$^{-8}$ & This work \\
		R4  & e$_{ph}$ + CO$_2$ & $\rightarrow$ O($^1$S) + CO & 
		Calculated &  This work \\
		R5  & e$_{ph}$ + CO & $\rightarrow$ O($^1$S) + C & 
		Calculated & This work \\
		R6  & e$_{ph}$ + O$_2$ & $\rightarrow$ O($^1$S) + O & 
		Calculated & This work \\
		R7  & e$_{ph}$ + O & $\rightarrow$ O($^1$S) + O & 
		Calculated & This work \\
		
		R8  & CO$_2^+$ + e$_{th}$  & $\rightarrow$ O($^1$S) + CO &
		4.2 $\times$ 10$^{-7}$ (300/T$_e$)$^{0.75}$  & 
		\cite{Viggiano05}\footnotemark[1]  \\         
		R9  & O$_2^+$ + e$_{th}$  & $\rightarrow$ O($^1$S) + CO &
		1.95 $\times$ 10$^{-7}$ (300/T$_e$)$^{0.7}$  & 
		\cite{Gkouvelis20}\footnotemark[3]   \\   
		R10  & h$\nu$ + CO$_2$ & $\rightarrow$ O($^1$D) + CO & 
		2.72 $\times$ 10$^{-7}$ & This  work \\
		R11  & h$\nu$ + CO & $\rightarrow$ O($^1$D) + C & 
		1.43 $\times$ 10$^{-8}$ & This  work \\
		R12  & h$\nu$ + O$_2$ & $\rightarrow$ O($^1$D) + O & 
		6.8 $\times$ 10$^{-7}$ & This work \\
		R13  & e$_{ph}$ + CO$_2$ & $\rightarrow$ O($^1$D) + CO & Calculated & 
		This work \\
		R14 & e$_{ph}$ + CO & $\rightarrow$ O($^1$D) + C & Calculated & This 
		work \\
		R15  & e$_{ph}$ + O$_2$ & $\rightarrow$ O($^1$D) + O & Calculated & 
		This work \\		
		R16 & e$_{ph}$ + O & $\rightarrow$ O($^1$D) + e & Calculated & 
		This work \\		
		R17  & CO$_2^+$ + e$_{th}$  & $\rightarrow$ O($^1$D) + CO &
		4.2 $\times$ 10$^{-7}$ (300/T$_{e}$)$^{0.75}$ &   
		\cite{Viggiano05}\footnotemark[2] \\ 
		R18  & CO$^+$ + e$_{th}$  & $\rightarrow$ O($^1$D) + CO &
		2.5 $\times$ 10$^{-8}$ (300/T$_{e}$)$^{0.55}$  & 
		\cite{Rosen98}\\       
		R19 & O$_2^+$ + e$_{th}$  & $\rightarrow$ O($^1$D) + O & 
		2.21 $\times$ 10$^{-7}$ (300/T$_{e}$)$^{0.46}$ & \cite{Guberman88} 
		\\   		    
		R20  & O($^1$S) + e$_{th}$ & $\rightarrow$  O($^1$D)  + e$_{th}$ & 
		8.5 $\times$ 10$^{-9}$  & \cite{Berrington81} 
		\\	    
		R21 & O($^1$S) + CO$_2$ & $\rightarrow$ O($^3$P) + CO$_2$ &
		3.21 $\times$ 10$^{-11}$ exp(-1323/Tn)  &          
		\cite{Capetanakis93} \\
		R22 & O($^1$S) + CO & $\rightarrow$ O($^3$P) + CO &
		7.4 $\times$ 10$^{-14}$ exp(-957/Tn) & \cite{Capetanakis93} \\
		R23 & O($^1$S) + O$_2$ & $\rightarrow$ O($^3$P) + O$_2$ & 
		2.32 $\times$ 10$^{-12}$ exp(-812/Tn) & \cite{Capetanakis93} \\	
		R24 & O($^1$S) + N$_2$ & $\rightarrow$ O($^3$P) + N$_2$ & 
		5 $\times$ 10$^{-17}$ & \cite{Atkinson72} \\           
		
		R25 & O($^1$S) + O($^3$P) & $\rightarrow$ 2 O($^1$D)  & 
		2 $\times$ 10$^{-14}$  & \cite{Krauss75} \\	    
		R26  & O($^1$S) + e$_{th}$ & $\rightarrow$  O($^3$P)  + e$_{th}$ & 
		7.3 $\times$ 10$^{-13}$ T$_e^{0.94}$ & \cite{Berrington81} \\	    
		
		R27  & O($^1$S) & $\rightarrow$ O($^1$D) + h{$\nu_{5577 \AA}$} &
		1.26 & \cite{Wiese96} \\   
		
		R28  & O($^1$S) & $\rightarrow$ O($^3$P) + h{$\nu_{2972 \AA}$} &
		0.075 & \cite{Wiese96} \\ 
		
		R29  & O($^1$D) + CO$_2$ & $\rightarrow$ O($^3$P) + CO$_2$ &
		6.8 $\times$ 10$^{-11}$ exp(117/T$_n$) & \cite{Streit76} \\
		R30 & O($^1$D) + N$_2$ & $\rightarrow$ O($^3$P) + N$_2$ & 
		1.8 $\times$ 10$^{-11}$ exp(107/T$_n$) & \cite{Atkinson97} \\
		R31  & O($^1$D) + CO & $\rightarrow$ O($^3$P) + CO & 
		1 $\times$ 10$^{-11}$ & \cite{Schofield78} \\
		R32 & O($^1$D) + O$_2$ & $\rightarrow$ O($^3$P) + O$_2$ & 
		3.2 $\times$ 10$^{-11}$ exp(67/T$_n$) & \cite{Atkinson97} \\	
		R33  & O($^1$D) + O($^3$P) & $\rightarrow$ 2 O($^3$P)  & 
		2.13 $\times$ 10$^{-12}$ + 2.60 $\times$ 10$^{-13}$ T$_n^{0.5}$  &   
		\\		
	    &  &  & 
		 - 	2.24 $\times$ 10$^{-15}$ T$_n$  &  \cite{Yee90} \\		    
		    
		R34 & O($^1$D) & $\rightarrow$ O($^3$P) + h{$\nu_{6300\AA}$} & 
		6.478 $\times$ 10$^{-3}$ & \cite{Froese04}\\
		R35 & O($^1$D) & $\rightarrow$ O($^3$P) + h{$\nu_{6364\AA}$} & 
		2.097 $\times$ 10$^{-3}$ & \cite{Froese04}\\
		
		R36 & O($^1$D) + e$_{th}$ & $\rightarrow$  O($^3$P) + e$_{th}$  & 
		1.6 $\times$ 10$^{-12}$ T$_e^{0.91}$  & \cite{Berrington81} 
		\\                			
		\hline
	\end{tabular}
	
	{The unattenuated photodissociative excitation frequencies for reactions 
	R1, R2, R3, R10, R11, \& R12 are calculated at heliocentric distance of 
	1.57 AU;{\footnotemark[1] This value is multiplied by 0.05, see 
	\cite{Gkouvelis20}}; \footnotemark[2] This rate coefficient is 
	multiplied with a factor 0.59, see \cite{Jain13b} and the main text; 
	\footnotemark[3] This rate coefficient is multiplied by a factor 0.09, 
	see \cite{Gkouvelis20} for more details; h$\nu$, e$_{ph}$, e$_{th}$, 
	T$_e$ and T$_n$ represent solar photon, photoelectron, thermal 
	electron, electron and neutral temperatures, respectively.}
\end{table*}

%%%%%%%%%%%%%%%%%%%%%%%%%%%%%%%%%%%%%%%%%%%%%%%%%%%%%%%%%%%%%%%%%%%
\section{Results and Discussion}
\label{sec:res_dis}
\begin{figure*}
	\centering
	\includegraphics[width=0.45\linewidth]{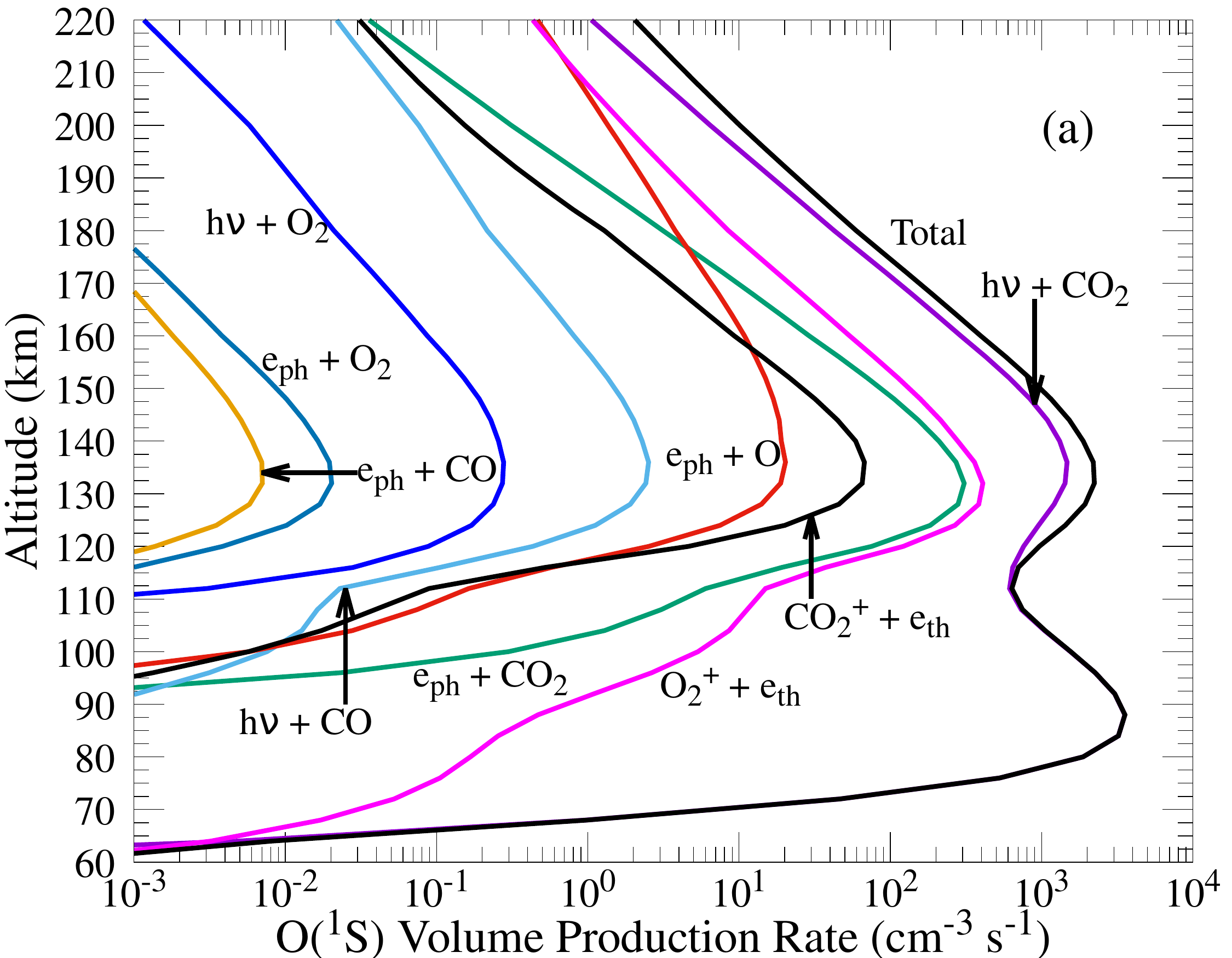}
	\includegraphics[width=0.45\linewidth]{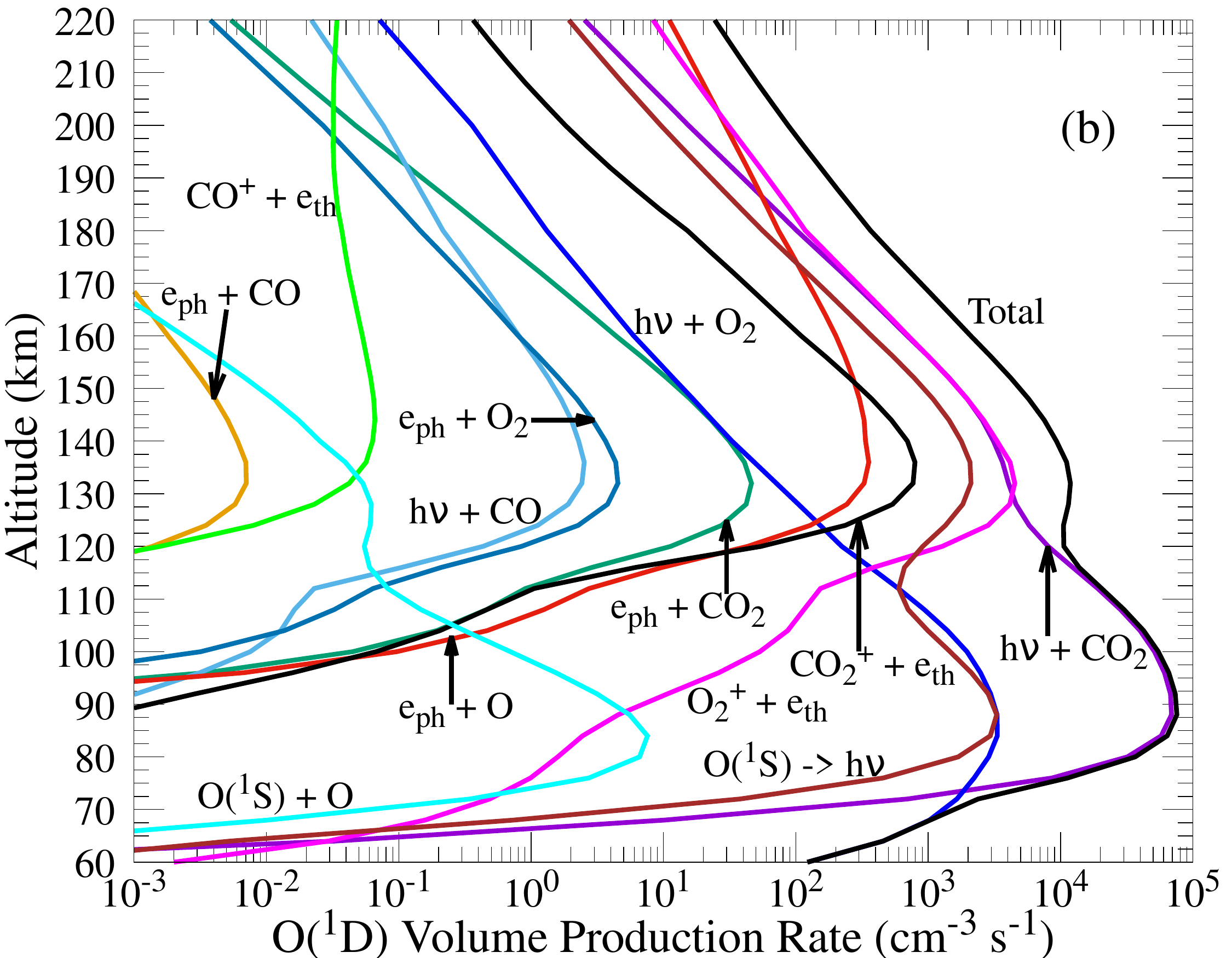}
	
	\includegraphics[width=0.45\linewidth]{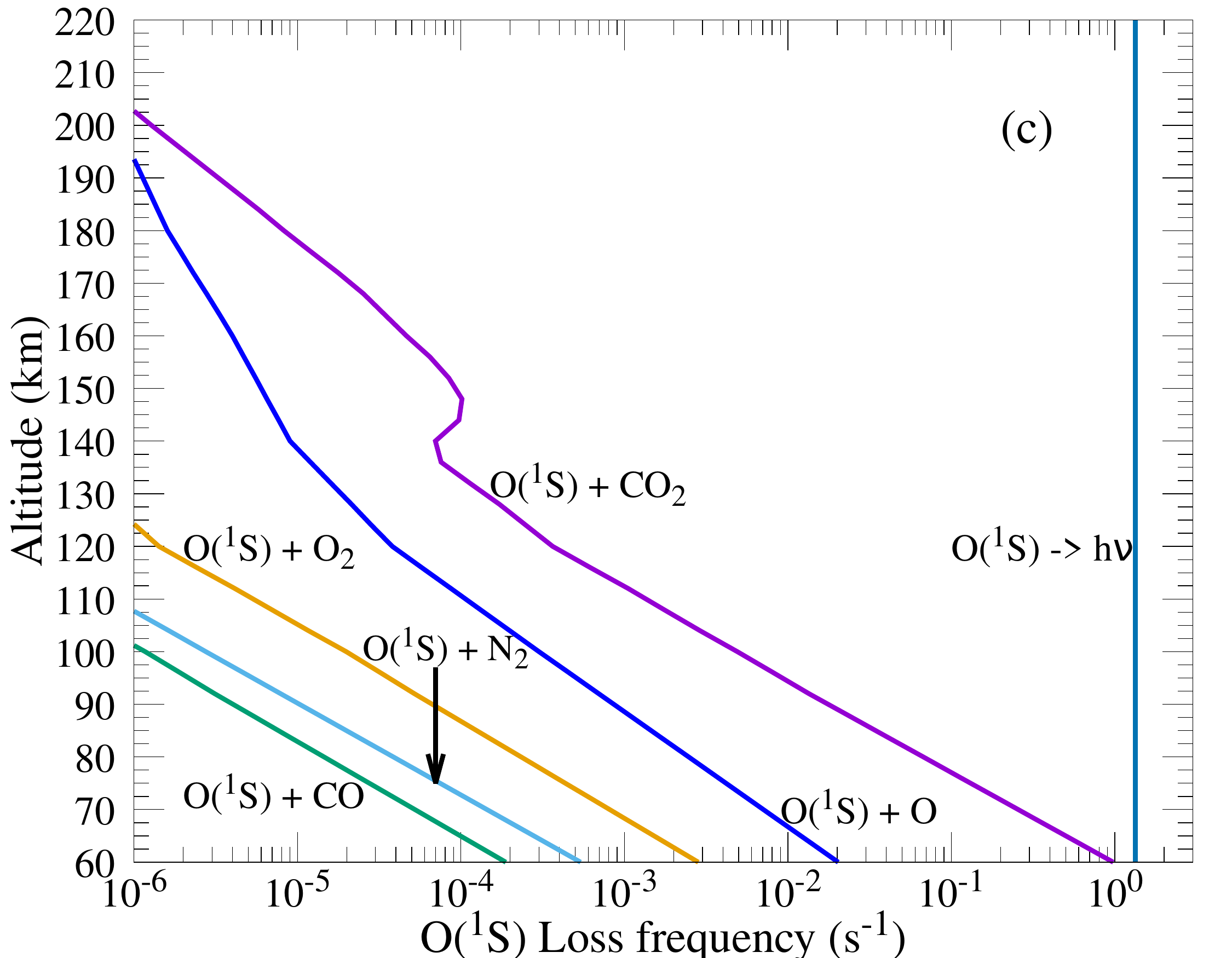}
	\includegraphics[width=0.45\linewidth]{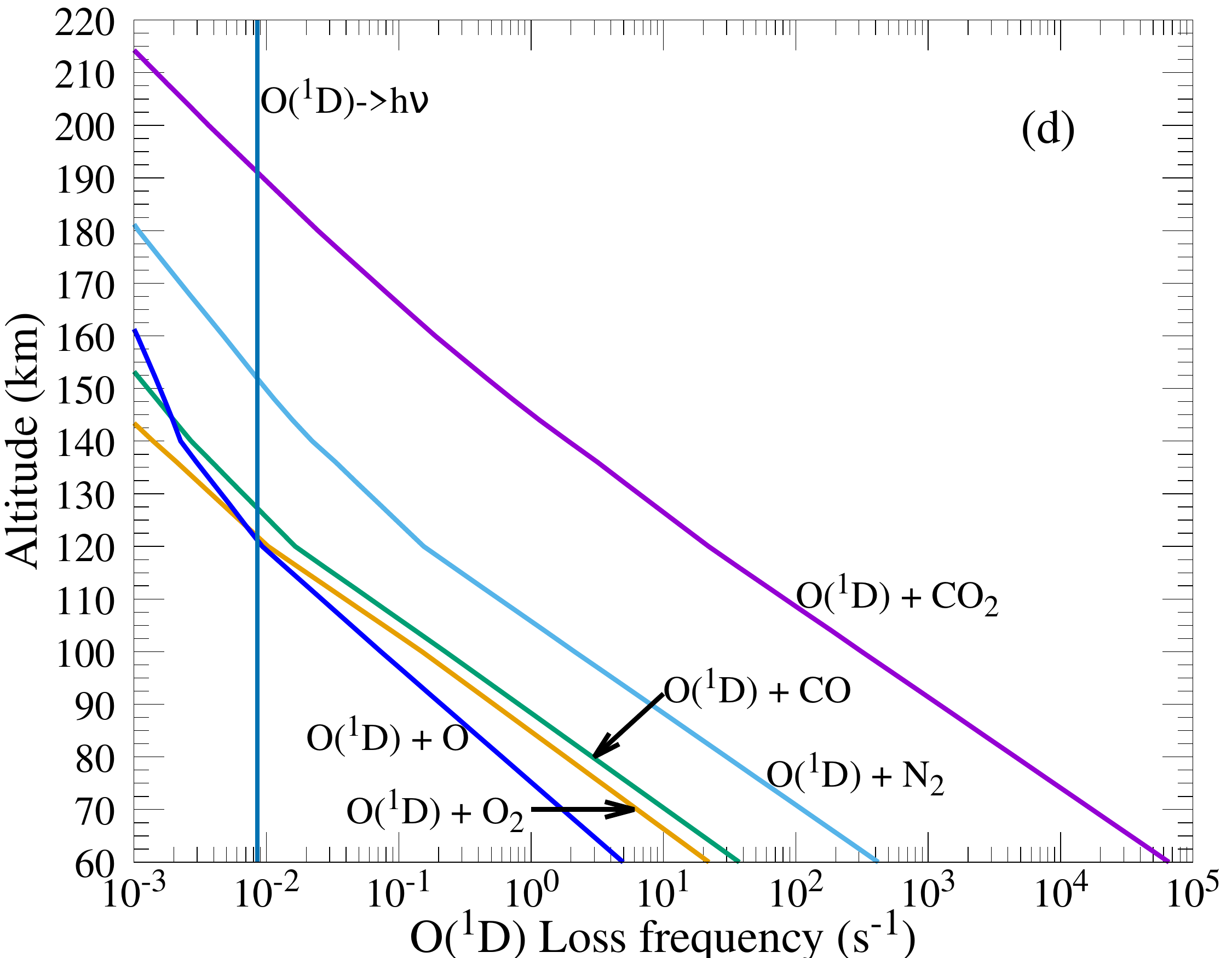}
	\caption{Modelled volume production rate  (top panels) and 
    loss frequency profiles (bottom panels) of O($^1$S) and O($^1$D) for  
	different photochemical processes in the Martian upper atmosphere 
	{for solar 
	zenith angle 30$^o$ using EUVM/MAVEN measured solar flux on 28 April 2019.}}
	\label{fig:o1sd_prlf}
\end{figure*}

The modelled production rate and loss frequency profiles of O($^1$S) and 
O($^1$D) for various photochemical processes in the Martian upper atmosphere 
are presented in Figure~\ref{fig:o1sd_prlf}. The calculated volume production 
rate profiles in Figure~\ref{fig:o1sd_prlf} (a) 
show that the total O($^1$S)  formation rate, which is majorly 
due to photodissociative excitation of CO$_2$,  has a double peak structure 
with lower and upper peaks at  altitudes of 85 and 135 km, respectively.   
Photodissociative excitation of CO$_2$ by the solar radiation flux at 
HI Lyman-$\alpha$ wavelength causes the lower production peak, whereas the 
upper 
production peak is  controlled by the flux mainly in wavelength  
region 860--1160 \AA. The  formation rate of O($^1$S) due to other excitation 
mechanisms is smaller by a factor of 3 or  more compared to that from 
photodissociative excitation of CO$_2$. 

Various modelled volume production rate profiles presented 
in Figure~\ref{fig:o1sd_prlf}  (b) show that the total formation rate of 
O($^1$D) is controlled by different excitation sources. The  lower volume 
production peak of O($^1$D) at  altitude of 90 km is mainly 
due to the photodissociative excitation of CO$_2$, whereas the upper peak (at 
an altitude of 130 km) is controlled by photodissociative excitation of CO$_2$, 
dissociative recombination of O$_2^+$, and radiative decay of O($^1$S). 
Electron impact excitation of atomic oxygen and dissociative recombination of 
O$_2^+$ contribute equally to the total O($^1$D) formation at  altitudes 
above 170 km. Four excitation sources, viz., photodissociative excitation  of 
CO$_2$, radiative decay of O($^1$S), electron impact on atomic oxygen, and 
dissociative recombination of O$_2^+$ together produce more than 95\% of total 
O($^1$D) in the altitude range of 60 to 220 km, whereas the other excitation 
sources play a minor role in the O($^1$D) formation. We also determined
the relative contributions of these excitation sources to the total formation 
of O($^1$D) and found that the photodissociative excitation of CO$_2$ and 
radiative decay of O($^1$S) together produce about 50\% of total O($^1$D) in 
the altitude range of 130 to 160 km and rest is majorly via electron impact on 
atomic oxygen and dissociative recombination of O$_2^+$. Above 170 km altitude, 
the 
electron impact on atomic oxygen and dissociative recombination of O$_2^+$ 
gradually takes over as major production sources of O($^1$D) (see 
Figure~\ref{fig:o1sd_prlf} (b)).

The modelled loss frequency profiles presented in Figure~\ref{fig:o1sd_prlf} 
(c) show that radiative decay is the dominant loss mechanism for O($^1$S)  in 
the Martian upper atmosphere. Owing to a short radiative lifetime (0.75 s), 
the collisional quenching  with Martian neutral species has no impact on  total 
loss frequency of  O($^1$S). As shown in Figure~\ref{fig:o1sd_prlf} (d), due  
to the long radiative lifetime ($\sim$120 s), the collisional quenching of 
O($^1$D) 
by CO$_2$ is significant at altitudes up to 190 km, and above this 
distance radiative decay takes over as the dominant loss mechanism. 

\begin{figure}
\centering
\includegraphics[width=\linewidth]{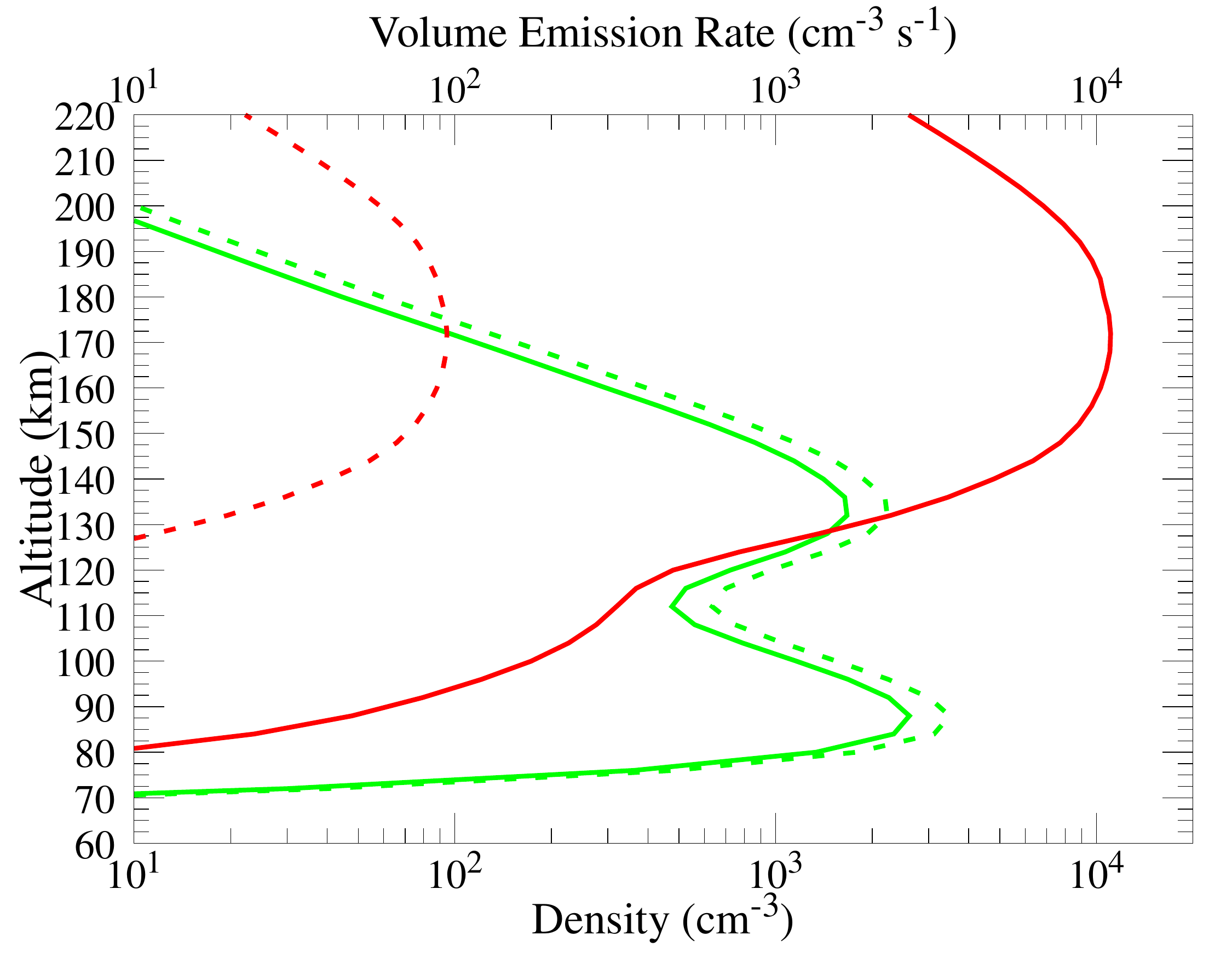}	
\caption{Modelled volume emission rate (dashed curves) and density (solid 
curves) profiles for O($^1$S) and O($^1$D) are plotted in the top and 
bottom x-axes, respectively. Green and red curves are the respective 
calculated profiles for O($^1$S) and O($^1$D).}
\label{fig:vemis_den}
\end{figure}

By incorporating the previously discussed different production and loss 
processes, the modelled density  and volume emission rate profiles for 
O($^1$S) and O($^1$D) are presented in Figure~\ref{fig:vemis_den}. The 
calculated density profile for O($^1$S) has a double peak structure with
respective upper and lower peaks at altitudes around 85 and 135 km (see 
solid green curve in Figure~\ref{fig:vemis_den}). These peaks are mainly 
due to the formation of O($^1$S) via photodissociative excitation of CO$_2$ 
by solar radiation flux in different wavelength regions (see 
Figure~\ref{fig:o1sd_prlf} (a)). The modelled O($^1$D) density profile has a 
broad peak in the altitude range of 150 to 200 km, which is determined by 
various production processes and the  collisional quenching of O($^1$D) by 
CO$_2$ (see solid red curve in Figure~\ref{fig:vemis_den}).  This calculation 
shows that strong collisional quenching of O($^1$D) substantially reduces its  
density at altitudes below 150 km. We calculated volume emission rates of these 
metastable species  by multiplying the modelled density profiles with 
corresponding transition probabilities  (see dashed curves in 
Figure~\ref{fig:vemis_den}). The radiative lifetime of O($^1$S) is smaller by 
more than two orders of magnitude compared to that of O($^1$D), which results 
in faster spontaneous decay  compared to that of  O($^1$D). Our 
modelled volume emission rates profiles show that radiative decay of O($^1$D) 
and O($^1$S)  
produce corresponding forbidden oxygen 
emissions with maximum intensities at  altitudes 
above and below 140 km  in the Martian upper atmosphere, respectively.

\begin{figure}
	\centering
	\includegraphics[width=\linewidth]{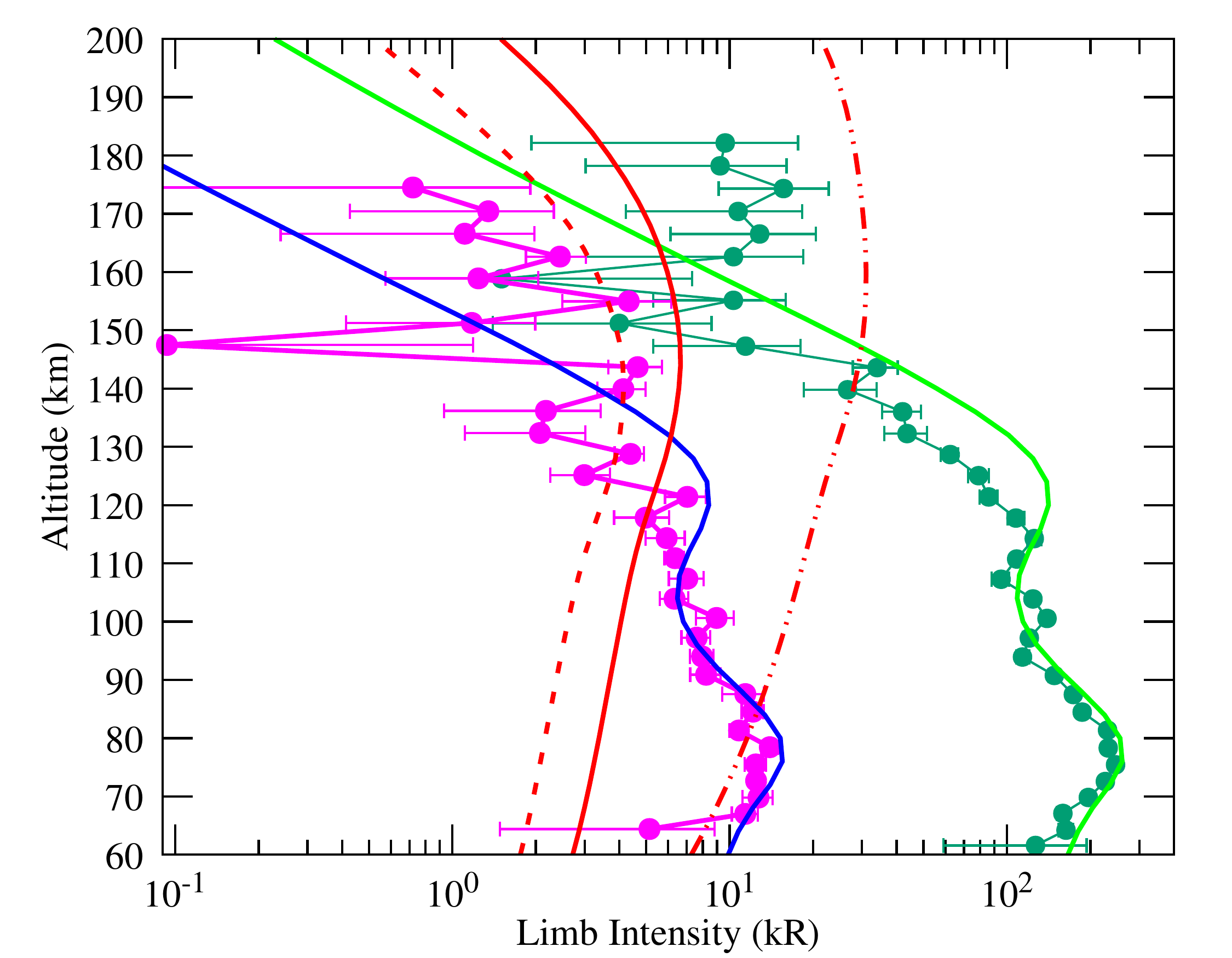}
	\caption{Comparison between modelled and observed limb intensity profiles 
		of [OI] 2972, 5577, and 6300 \AA\ emissions. Magenta and thick green 
        curves with x-error bars represent the respective NOMAD-TGO 
		observed limb intensity profiles for [OI] 2972 and  5577 \AA\ emissions 
        on 28 April 2019 \cite[taken from][]{Gerard20}. Blue, green and red 
        curves are the modelled limb intensity profiles for [OI] 2972, 5577 and 
        6300 \AA\ emissions, respectively. These limb intensities are 
        calculated by decreasing the \cite{Fox04} modelled neutral density 
        profiles for solar minimum condition
        by a factor of 2.
		Dashed  and dash double-dotted  red curves represent the calculated 
		limb intensity profiles for [OI] 6300 \AA\ emission  by reducing the 
        \cite{Fox04} modelled atomic oxygen density 
		by a factor of 10 and for solar maximum condition, respectively.}
	\label{fig:limb_inten}
\end{figure}

{A comparison between the modelled limb emission intensity profiles and 
NOMAD-TGO observations  for [OI] 2972 and  5577 \AA\ emissions is presented in 
Figure~\ref{fig:limb_inten}.} By decreasing the 
neutral density profiles of \cite{Fox04} model by a factor of {2}, which 
we call a standard case, the 
calculated limb emission intensity profiles for [OI] 2972 and 5577 \AA\ 
emissions are found to be consistent with NOMAD-TGO observations at altitudes 
below 150 km (see solid blue and green curves in Figure~\ref{fig:limb_inten}). 
This agreement suggests that photodissociative excitation of CO$_2$ is 
sufficient to explain the NOMAD-TGO observed limb intensities for [OI] 2972 and 
5577 \AA\ emissions. At altitudes above 140 km, our modelled limb 
intensities are decreasing rapidly and  smaller  by an order of magnitude 
compared to  NOMAD-TGO observation. However, it should be noted that 
the uncertainties in the  NOMAD-TGO measured limb intensities for [OI] 2972 and 
5577 \AA\ emissions are larger at altitudes above 140 km compared to those 
at lower altitudes. 

{We have considered \cite{Fox04} modelled neutral density profiles, 
which 
are 
based on the Viking measurements and for solar minimum condition, to calculate 
the limb intensities of forbidden atomic oxygen emission lines. As shown in 
Figure~\ref{fig:neudens}, the in-situ measured CO$_2$ and O densities vary 
about an order of 
magnitude at altitudes around 200 km. Considering this variability into 
account, 
a reduction in the neutral density of \cite{Fox04} model by a factor of 2 is 
necessary to explain the observed limb intensity profiles. We compared the 
\cite{Fox04} neutral density profiles for solar minimum condition with Mars 
Climate Database (MCD) modelled neutral densities for NOMAD-TGO observational 
condition and found that they are nearly consistent. By reducing the neutral 
density profiles of MCD by a factor of 2, \cite{Gerard20} also  
explained the observed emission limb intensity profiles of [OI] 2972 and 5577 
\AA\ emissions. Our calculations and 
\cite{Gerard20} study showed that 
the observed limb intensity profiles of forbidden oxygen emission (5577 \& 2972 
\AA) lines can be 
used to constrain the neutral abundances, particularly CO$_2$, in the Martian 
upper atmosphere. }

%{Our input solar flux is from Flare Irradiance Spectral Model-Mars 
%(FISM-M) 
%model developed by \cite{Thiemann17}, which is based on in-situ measured solar 
%fluxes by Extreme Ultraviolet Monitor instrument onboard Maven mission in the 
%0-7, 17-22, and 117-125 nm wavelength regions \citep{Eparvier15}. 
%\cite{Thiemann17} have studied 
%the 
%uncertainties associated with the daily modelled flux at different wavelength 
%regions. The dissociative 
%excitation of CO$_2$, which produces [OI] 2972 and 5577 \AA\
%emissions, mainly depends on the input solar flux at HI Ly-$\alpha$ 
%wavelength and in 
%the 
%wavelength region 860-1160 \AA. Hence, the uncertainties associated with 
%FISM-M 
%modelled flux in these wavelength regions directly can impact our modelled 
%limb 
%intensities. When we reduce the input solar flux by 35\%, the modelled limb 
%intensity profiles of [OI] 2972 and 5577 A emissions are consistent with the 
%NOMAD-TGO observation. This agreement suggests that the reduction in the input 
%solar 
%flux is necessary to explain the observed limb intensities of forbidden oxygen 
%emission lines in the Martian upper atmosphere.}

\begin{figure*}
	\centering
	\includegraphics[width=0.47\linewidth]{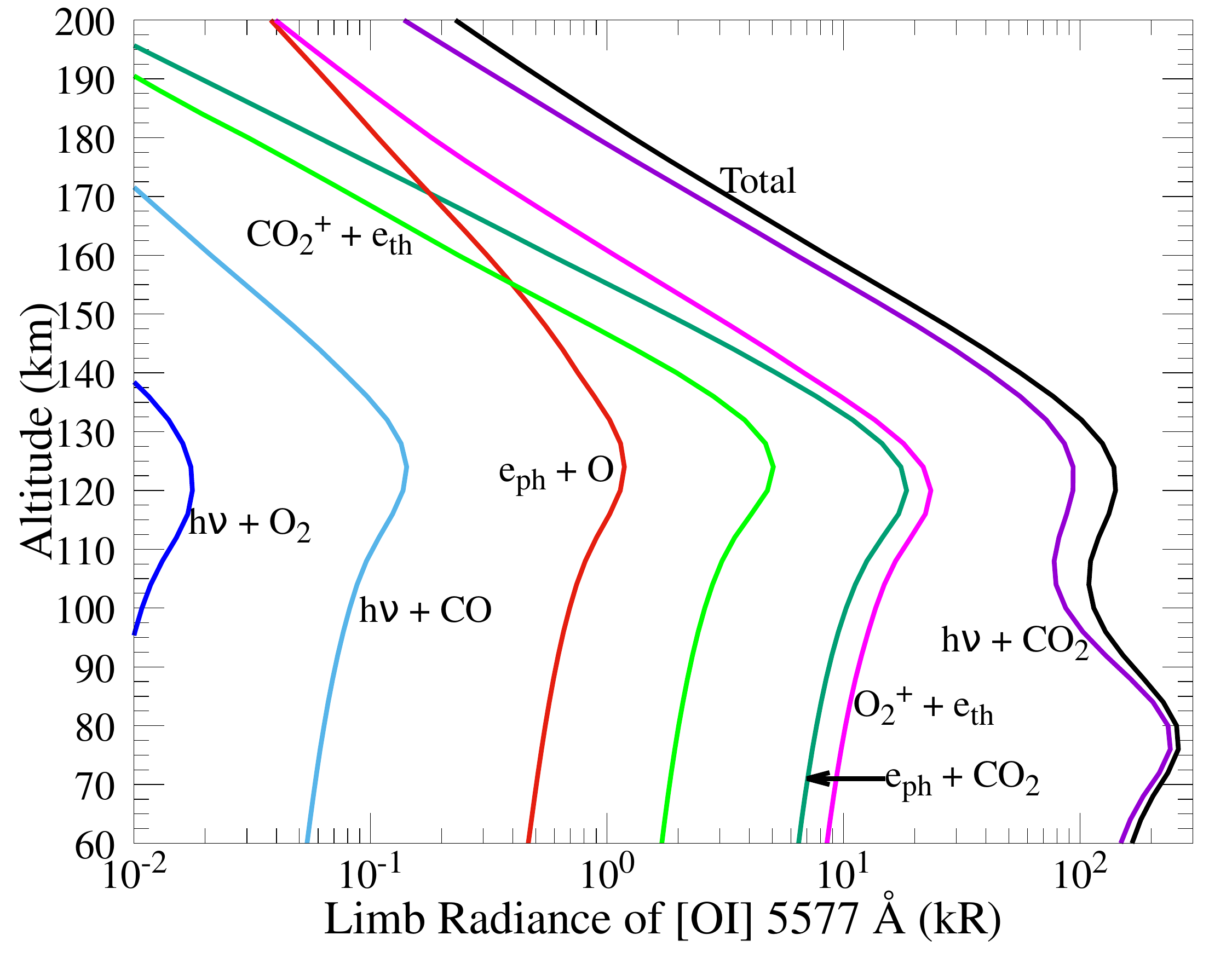}
	\includegraphics[width=0.47\linewidth]{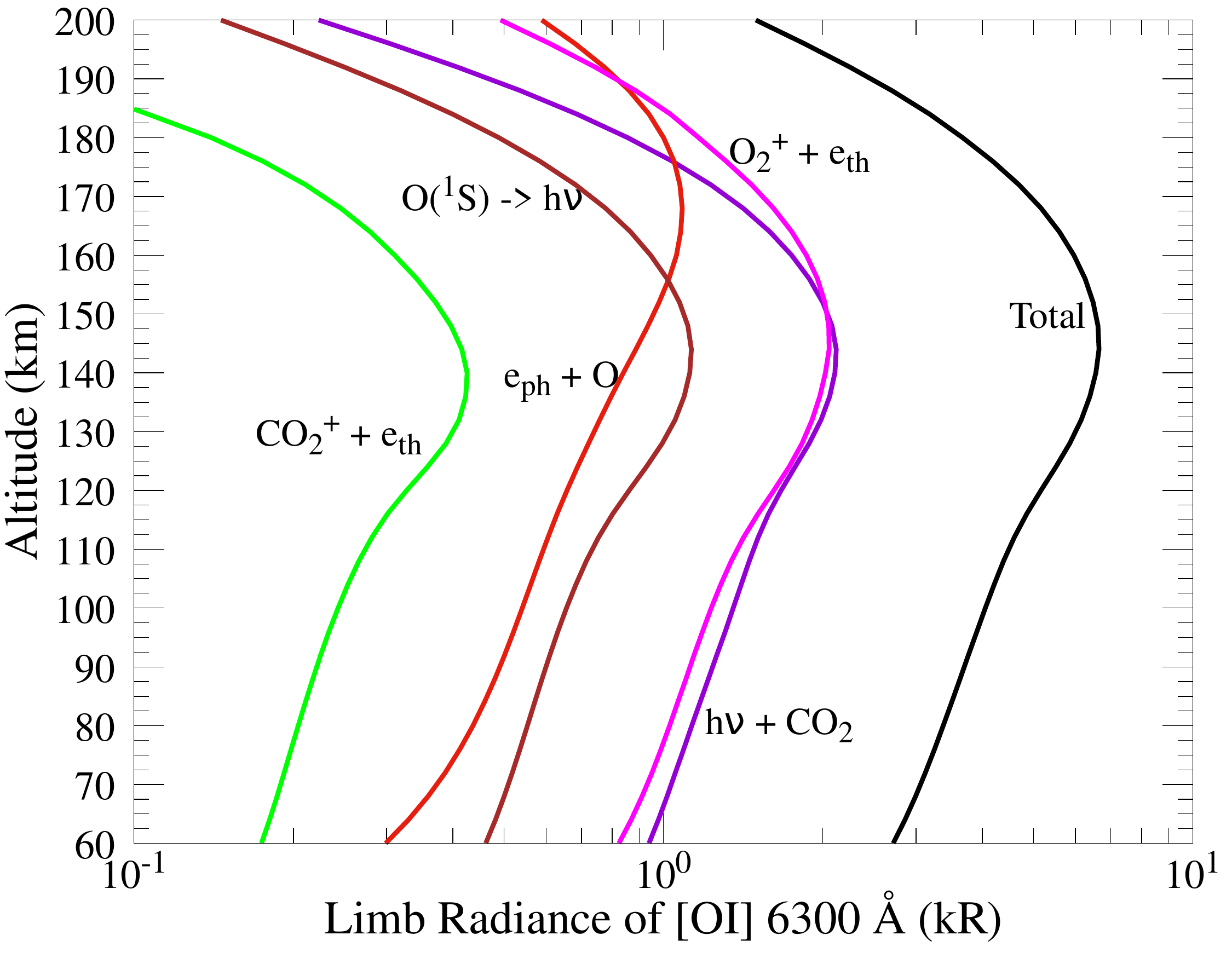}
	
	\caption{Modelled limb intensity profiles of [OI] 5577  \AA\ (left panel) 
		and 
		6300 \AA\ (right panel) emissions for major formation processes in the 
		Martian upper atmosphere.}
	\label{fig:red-green-emissions}
\end{figure*}

{For our standard case, the modelled intensity profiles of [OI] 5577 and 
6300 \AA\ emissions  via major production processes  are presented in  
Figure~\ref{fig:red-green-emissions}. The calculated limb intensity profiles in 
the left panel of this figure show that most of the [OI] 5577 \AA\ emission 
produces via photodissociative excitation of CO$_2$  in the limb viewing 
geometry with a contribution more than 80\%  in the altitude range 
60 to 200 km. The remaining significant contributions are from the thermal 
recombination of O$_2^+$ and electron 
impact excitation of atomic oxygen and CO$_2$.  Thus, the 
observed limb intensity of this emission can be used 
to constrain the CO$_2$ density in the upper atmosphere of Mars.}

{In the case of [OI] 6300 \AA\ emission, for the altitudes below 170 km, 
we determined that the total limb 
intensity is controlled by  photodissociation of CO$_2$, thermal recombination 
of O$_2^+$, electron impact on atomic oxygen, and radiative decay of O($^1$S),
with a relative contribution of about 35\%, 30\%, 10\% and   15\%, respectively
(see the right panel of 
Figure~\ref{fig:red-green-emissions}). The remaining 10\% of  emission 
intensity is determined by several excitation sources as shown in 
Figure~\ref{fig:o1sd_prlf} (b). It can be noticed in this figure that above 170 
km altitude, electron impact on 
atomic oxygen and thermal recombination of O$_2^+$ 
significantly contribute to the total limb intensity of [OI] 6300 \AA\ 
emission. In the following section we discuss 
the role of various parameters which can influence the observed [OI] 
red-doublet 
limb intensities in the Martian upper atmosphere.}

\subsection{Effect of input parameters on the modelled red-doublet limb 
	intensities}
The branching ratios for O($^1$D) radiative decay transitions show that the 
emission intensity 
of [OI]  6364 \AA\ line should be about one third that of [OI]  6300 \AA\ 
\citep[see emission rates for reactions R34 and R35 in 
Table~\ref{Tab:reactions} and][]{Wiese96}. Hence, we present the calculated 
limb intensity profile only for [OI] 6300 \AA\ emission  and 
compare its magnitude with those of [OI] 2972 and  5577 \AA\ emissions in 
Figure~\ref{fig:limb_inten} (see red curves in this Figure).

The modelled limb 
intensity profile for [OI] 6300 \AA\ emission, which is calculated  for our 
standard case, has a broad peak in the altitude range of 120 to 170  km  and 
comparable with the upper limit of [OI] 2972 \AA\ observation (see solid red 
curve in Figure~\ref{fig:limb_inten}). At  altitudes below 120 km, the 
modelled limb intensity of [OI] 6300 \AA\ emission is smaller by a factor of 
2 to 5 compared to that of NOMAD-TGO observation for [OI] 2972 
\AA\ emission (see solid red and magenta curves in 
Figure~\ref{fig:limb_inten}).  This calculation shows that the modelled 
limb intensity for [OI] 6300 \AA\ emission is significant compared to that of 
[OI] 2972 \AA\ and should be observable in NOMAG-TGO dayside spectra  
{taken in 
the altitude region of 120 to 170 km,} provided 
sufficient signal-to-noise ratio in the measurement during the observation 
period. 

%{As discussed before, the observed [OI] red-doublet emission intensity 
%in the altitude range 60 to 200 km is significantly controlled by 
%photodissociative excitation of CO$_2$, dissociative recombination of O$_2^+$, 
%electron impact on atomic oxygen, and  radiative decay of O($^1$S) (see right 
%panel of Figure~\ref{fig:red-green-emissions}).}    

As pointed out by \cite{Huestis06}, the photodissociative cross section of 
CO$_2$ producing O($^1$D) was not experimentally determined before. But there 
are several recent developments in measuring the cross section for this 
excitation processes \citep{Sutradhar17,Lu15,Song14,Gao19}. However, these 
measured cross sections are limited to a small wavelength range from threshold 
to  1000 \AA. Recently, \cite{Raghuram20b} modelled atomic oxygen green and 
red-doublet 
emissions in CO-dominated and water-poor comet C/2016 R2 (Pan-STARRS). By 
accounting for various photochemical processes, 
{it was} found that both green and red-doublet atomic oxygen 
emissions are controlled by photodissociative excitation of CO$_2$ in the 
cometary coma of this comet.  They also showed that the uncertainties 
associated with the photon cross section of CO$_2$ producing O($^1$D) play an 
important role in determining the observed atomic oxygen green to red-doublet 
emission intensity ratio in comet C/2016 R2 (Pan-STARRS). By comparing the 
modelled and observed atomic oxygen green to red-doublet emission intensity 
ratios,  \cite{Raghuram20b} suggested that the photodissociative excitation 
cross section of CO$_2$ producing O($^1$D) should be increased by a 
factor of 3. Considering the uncertainty associated with this excitation 
process, when we increase the cross section for photodissociative excitation of 
CO$_2$ by a factor of 3, the modelled total [OI] 6300 \AA\ limb intensity is 
found to be  60\% higher compared to that of our standard case. This  
calculation suggests that the role of uncertainty associated with  
photodissociative  excitation of CO$_2$ lead to higher [OI] red-doublet 
limb intensities than our standard case.

Due to lack of experimentally determined value, following the assumption of 
\cite{Jain13b}, we incorporated branching ratio for dissociative recombination 
of CO$_2^+$ producing O($^1$D) as 0.59. Our calculations in 
Figure~\ref{fig:o1sd_prlf} (b) show that  the dissociative recombination of 
CO$_2^+$ plays a minor role in the total production  of O($^1$D). Even with the 
assumed higher branching ratio, we find that this excitation process 
contributes little {($<$10\%)} to the total formation of O($^1$D) when 
compared to that from other excitation sources in the Martian upper atmosphere. 
Hence, our assumed branching ratio has no impact on the modelled [OI] 
red-doublet limb 
intensities. Similarly, considering branching ratio from \cite{Gkouvelis18} 
for dissociative recombination of CO$_2^+$, our modelled production rate of 
O($^1$S)  is smaller by an order of magnitude or more compared to 
that from photodissociation of CO$_2$. Thus, the assumed  branching ratio for 
dissociative recombination of CO$_2^+$ has no impact on the modelled [OI] 2972 
and 5577 \AA\ limb intensities.

Collisional quenching of O($^1$D) has an important role in  determining the  
limb intensities of [OI] red-doublet emissions in the Martian upper atmosphere.
{By comparing the modelled total production rate and loss frequency 
profiles in 
Figure~\ref{fig:o1sd_prlf} we can understand that though O($^1$D) is produced 
a by an order magnitude higher than that of O($^1$S) for the altitudes 
below 140 km,} the strong collisional 
quenching  by CO$_2$ 
substantially reduces its 
density {at lower altitudes} (see Figure~\ref{fig:vemis_den}). We 
evaluated that about 30\% of the total O($^1$D) production is due to 
photodissociative excitation of CO$_2$ at altitudes above 140 km, and remaining 
is due to other excitation 
sources. Hence, larger amount of CO$_2$ significantly leads to strong 
collisional quenching of O($^1$D) rather than its production. But it should be 
noted that the large amount CO$_2$ density in the Martian upper atmosphere also 
leads to {an} increase in the [OI] 2972 and 5577 \AA\ limb emission 
intensities.  
The compilation of various experimentally determined O($^1$D) quenching rates 
by \cite{Burkholder15} and recent measurements by \cite{Reyes18}  showed that 
the uncertainty in measuring the rate coefficient of collisional reaction 
between CO$_2$ and O($^1$D) is  only about 20\%, which suggests that larger 
quenching of O($^1$D) in the Martian upper atmosphere is unlikely.

As discussed earlier, at altitudes above 170 km, the contribution 
from dissociative recombination of O$_2^+$ and electron impact excitation of 
atomic oxygen  is significant to the total formation of O($^1$D) (see 
Figure~\ref{fig:o1sd_prlf} and {also the right panel of 
Figure~\ref{fig:red-green-emissions}}). The formation of O$_2^+$ ion is mainly 
due the 
collisions between CO$_2^+$ and atomic oxygen. 
Hence, the change in neutral density profile of atomic oxygen can influence 
 O$_2^+$ ion density and also total volume production of O($^1$D) in the 
Martian upper atmosphere. As shown in Figure~\ref{fig:neudens}, the  
NGIMS/MAVEN measured atomic oxygen and CO$_2$ densities during September 2019 
are varying by an order of magnitude at  altitudes  above  170 km. Using 
NGIMS/MAVEN measurements, we find that the measured volume mixing ratios of 
O/CO$_2$  are smaller by a factor of 2 to 10  compared to that of \cite{Fox04}  
in the altitude range of 150 to 200 km. Considering the 
variability in  NGIMS/MAVEN 
measured densities into account, we decreased the \cite{Fox04} atomic oxygen 
density by an order of magnitude  to study its  impact on the modelled [OI] 
6300  \AA\ limb intensity. By decreasing the atomic oxygen density, we found 
that the modelled [OI] 6300 \AA\ limb intensity is smaller by maximum factor of 
2 compared to our standard case (see dashed red-curve in 
Figure~\ref{fig:limb_inten}). In this case, the contributions from both 
electron impact on atomic oxygen and thermal recombination of O$_2^+$ to the 
total 
O($^1$D) production significantly decrease in {the altitude range 60 to 
200 km}, whereas the radiative decay of O($^1$S) and 
photodissociation of CO$_2$ majorly 
produce O($^1$D) and cause the [OI] red-doublet emissions (see the right panel 
of 
Figure~\ref{fig:red-green-emissions}).  This calculation 
shows that the variation in  atomic oxygen density in the Martian upper 
atmosphere can lead to a significant change in the observed limb intensities of 
[OI] red-doublet emissions.

The  NGIMS/MAVEN measurements in the altitude range of 
150 to 200 km  shows that the CO$_2$ and O 
densities can vary by an order 
magnitude  {and are also smaller} than those of \cite{Fox04} model (see 
Figure~\ref{fig:neudens}). When we scaled the neutral densities of \cite{Fox04} 
model to the lower limit of NGIMS/MAVEN measurements, the  
calculated limb intensity for [OI] 6300 \AA\ is closer to our standard case at  
altitudes below 120 km. This calculation shows that in spite of lower 
neutral densities than those of \cite{Fox04},  the observed [OI] 6300 
\AA\ is not expected to change significantly compared to our standard case
at altitudes below 120 km. Thus, the modelled limb intensity profile of [OI] 
6300 \AA\ for our standard 
case serves as an upper limit during NOMAD-TGO observation period.

The thermal structure of Martian upper atmosphere  also plays an 
important role in determining the neutral densities and subsequently the 
emission intensities of forbidden atomic oxygen emission lines. During  
solar active condition, the densities of atomic oxygen and CO$_2$ 
increase in the upper atmosphere of Mars. We noticed in our calculations that  
the larger volume mixing ratio of atomic oxygen leads to higher limb 
intensities of [OI] 
2972 and 5577 \AA\ emissions only for altitudes above 150 km. But when we 
increase the atomic oxygen density in the model, the calculated limb intensity 
for [OI] 6300 \AA\ emission also increased in the altitude range of 60 to 200 
km. Hence, for higher thermospheric temperature all these emission lines are 
expected to  be observed  in the NOMAG-TGO dayside spectra. However, it 
should be noted that the 
NOMAD-TGO observations were carried out on  Mars 
for solar longitudes between 16$^\circ$ and 115$^\circ$ when it was crossing 
the aphelion and the solar activity during this period was also low. Thus,
larger neutral densities are not expected in the Martian upper 
atmosphere during NOMAD-TGO observation period.

During the NOMAD-TGO 
observation period, the background intensity level around  wavelength
6300 \AA\  is equivalent to more than 10 kR. Moreover, the [OI] red-doublet 
emission lines are closer to the longer wavelength limit of the NOMAD-TGO 
instrument and the detector has a 
reduced sensitivity in 
this region, which is the main reason for the absence of these emissions in the
observed dayside spectra \cite[personal communication,][]{Gerard20}. 
{Furthermore,} most of the 
NOMAD-TGO observations took place  when Mars was moving away from the Sun. But 
when Mars reaches perihelion, due to proximity of the Sun, the 
neutral densities of Martian upper atmosphere increases due to large 
thermospheric temperature. We evaluated the  limb intensity 
of [OI] 6300 \AA\ emission during high solar activity and when Mars is at 
perihelion. For this case,  we accounted for neutral atmosphere from 
\cite{Fox04} for solar maximum condition and used the EUVM/MAVEN 
measured solar flux on 12 December 2014 
(during which Mars was at perihelion and solar activity was also high, 
F$_{10.7}$ = 152 $\times$ 10$^{-22}$ W m$^{-2}$ Hz$^{-1}$).  We find that the  
modelled limb intensity [OI] 6300 \AA\ emission is higher by a more than a 
factor of 3 than our standard case (see dash double-dotted red curve in 
Figure~\ref{fig:limb_inten}). {In this case, the modelled 
[OI] 6300 \AA\ peak emission limb intensity is about 30 kR at altitudes above 
140 km and this emission line would have be seen in NOMAD-TGO spectra, provided 
that the 
background intensity is smaller compared to the modelled limb emission 
intensity.} 
Thus, we suggest that the simultaneous detection of 
[OI] red-doublet emission along with other atomic forbidden emissions on the 
dayside Martian upper atmosphere by NOMAD-TGO would be possible when Mars is at 
perihelion.

Our calculations showed that there are a few parameters which determine the 
limb 
intensities of [OI] red-doublet emissions in the Martian upper atmosphere. In 
spite of variability in the neutral 
densities in the Martian upper atmosphere, we find that the modelled limb 
intensity of [OI] 6300 \AA\ emission  should be smaller than that of [OI] 2972 
\AA\ emission by a factor of 2 to 5 at altitudes below  120 km and it is 
higher above this altitude (see 
Figure~\ref{fig:limb_inten}).  NOMAD-TGO should be able to detect these 
emissions 
in the 
dayside spectra provided sufficient signal-to-noise ratio during the 
observation.  Instrument sensitivity coupled with 
uncertainties in the neutral 
atmosphere (given the large  variability in atomic oxygen and CO$_2$ as shown 
Figure~\ref{fig:neudens}) may explain the absence of [OI] red-doublet emission 
lines in the dayside spectra taken during NOMAD-TGO observation period. 
However, 
when Mars gets closer to the 
Sun, the seasonal and solar cycle variations significantly affect the neutral 
densities of Martian upper atmosphere and result in the [OI] 6300 
\AA\ limb emission intensity within the observation limit of NOMAD-TGO.
More simultaneous observations of forbidden atomic  oxygen emissions are 
required to study the photochemistry of these emissions in the Martian upper 
atmosphere.

\section{Summary and Conclusions}
\label{sec:sum_con}
Recently, Nadir and Occultation for Mars Discovery ultraviolet and visible 
spectrometer instrument on board the European Space Agency’s ExoMars Trace Gas 
Orbiter made simultaneous detection of forbidden atomic oxygen emissions 
at wavelengths 2972 and 5577 \AA\ in the spectra observed between 24 April and 
1 December 2019 on the dayside Martian upper atmosphere. Thanks to the wide 
detection wavelength range of this instrument, it can measure the limb emission 
intensities of four forbidden atomic oxygen emissions at wavelengths 2972, 
5577, 6300, and 6364 \AA\ simultaneously. Since {both}  [OI] 2972 
and 
5577 \AA\  
emissions originate from the same excited state of atomic oxygen, detection of 
any one of these  emissions confirms the presence of other. Similarly, 
detection of [OI] 5577 \AA\ emission line also indicates the presence of [OI] 
red-doublet emissions at wavelengths 6300 and 6364 \AA, but the opposite is not 
true. However, these [OI] red-doublet emissions were not observed in the 
{visible} spectra taken during the NOMAD-TGO observation period due to 
the 
reduced sensitivity of the detector. By  accounting for the important chemical 
pathways of O($^1$S) and O($^1$D), which are the excited states of these 
forbidden atomic oxygen emissions, we developed a model to study the 
photochemistry of these forbidden emissions in the Martian upper atmosphere and 
aimed to explore the suitable conditions to observe [OI] red-doublet emissions 
along with other atomic oxygen forbidden emissions. Our  calculations show that 
NOMAD-TGO observed limb intensities for [OI] 2972 and 5577 \AA\  emissions are 
mainly  controlled by photodissociative excitation of CO$_2$. On 
reducing neutral density profiles of 
\cite{Fox04} model for solar minimum condition by a factor of {2}, our 
modelled limb 
intensities of  [OI] 2972 and 5577 \AA\ emissions are found to be in agreement 
with  NOMAD-TGO observations. 

We studied the role of different parameters which can influence the observation 
of  [OI] red-doublet emissions  in the dayside Martian upper 
atmosphere.  The  limb emission intensities of these emissions are found to be  
controlled by 
photodissociative excitation of CO$_2$, radiative decay of O($^1$S), 
dissociative recombination of O$_2^+$, and electron impact on atomic oxygen. 
Our modelled  limb 
intensity profile of [OI] 6300 \AA\ emission 
is comparable and higher than that of NOMAD-TGO observation for [OI] 2972 \AA\ 
emission in the altitude range of 60 to 200 km. We find that the peak limb 
emission intensity 
for [OI] 6300 \AA\ emission occurs at altitudes above 120 km and  higher 
than the
upper limit of NOMAD-TGO observation for [OI] 2972 \AA\ emission. But at 
altitudes below 120 
km,  the 
modelled limb intensity for [OI] 6300 \AA\ emission is smaller by a factor of 2 
to 5 compared to that of NOMAD-TGO observation for [OI] 
2972 \AA\ emission. Hence,   [OI] 6300 
\AA\ 
emission line 
is expected to be observed in 
the dayside spectra of NOMAD-TGO at altitudes {above 140 km} 
{provided 
that} the 
signal-to-noise ratio is sufficient during the observation period.  
Due to the reduced detector sensitivity around the wavelength 6300 \AA, 
NOMAD-TGO could not observe the [OI] red-doublet emission lines in the dayside 
of Mars. Moreover, most of 
the NOMAD-TGO observations took place when the Mars was travelling away from 
the Sun (L$_s$  is varying from 
16$^\circ$ to 115$^\circ$), during which the neutral densities in the 
Martian upper atmosphere are smaller compared to those at near perihelion. 
Based on 
our modelling we suggest that all these atomic forbidden 
emissions should be observable in the dayside NOMAD-TGO spectra taken over the 
southern hemisphere, when Mars is at perihelion.  More simultaneous 
observations of forbidden atomic oxygen 
emission lines are required to study the photochemistry of Martian upper 
atmosphere.

\section*{Acknowledgements}
SR is supported by Department of Science 
and Technology (DST) with Innovation in Science Pursuit for 
Inspired Research (INSPIRE) faculty award 
[grant : dst/inspire/04/2016/002687], and he would like to 
thank Physical Research Laboratory for facilitating 
conducive research environment. {Authors would like to thank the 
anonymous reviewers for their constructive comments and suggestions that 
improved this manuscript.}

\section*{Data Availability}
This paper make use of NGIMS/MAVEN measured 
neutral and ion number densities L2 data  which has been accessed through the 
web link (\href{https://pds-atmospheres.nmsu.edu} 
{https://pds-atmospheres.nmsu.edu})  The modelled data 
in this research will be shared on reasonable request
to the corresponding author.


\bibliographystyle{cas-model2-names}

%\bibliography{cas-refs}
%\bibliography{References/agusample} % if your bibtex file is called example.bib
%\bibliography{References/references-raghu-abr}

\begin{thebibliography}{42}
	\expandafter\ifx\csname natexlab\endcsname\relax\def\natexlab#1{#1}\fi
	\providecommand{\url}[1]{\texttt{#1}}
	\providecommand{\href}[2]{#2}
	\providecommand{\path}[1]{#1}
	\providecommand{\DOIprefix}{doi:}
	\providecommand{\ArXivprefix}{arXiv:}
	\providecommand{\URLprefix}{URL: }
	\providecommand{\Pubmedprefix}{pmid:}
	\providecommand{\doi}[1]{\href{http://dx.doi.org/#1}{\path{#1}}}
	\providecommand{\Pubmed}[1]{\href{pmid:#1}{\path{#1}}}
	\providecommand{\bibinfo}[2]{#2}
	\ifx\xfnm\relax \def\xfnm[#1]{\unskip,\space#1}\fi
	%Type = Article
	\bibitem[{{Atkinson} et~al.(1997){Atkinson}, {Baulch}, {Cox}, {Hampson},
		{Kerr}, {Rossi} and {Troe}}]{Atkinson97}
	\bibinfo{author}{{Atkinson}, R.}, \bibinfo{author}{{Baulch}, D.L.},
	\bibinfo{author}{{Cox}, R.A.}, \bibinfo{author}{{Hampson}, Jr., R.F.},
	\bibinfo{author}{{Kerr}, J.A.}, \bibinfo{author}{{Rossi}, M.J.},
	\bibinfo{author}{{Troe}, J.}, \bibinfo{year}{1997}.
	\newblock \bibinfo{title}{{Evaluated Kinetic and Photochemical Data for
			Atmospheric Chemistry: Supplement VI. IUPAC Subcommittee on Gas 
			Kinetic Data
			Evaluation for Atmospheric Chemistry}}.
	\newblock \bibinfo{journal}{Journal of Physical and Chemical Reference Data}
	\bibinfo{volume}{26}, \bibinfo{pages}{1329--1499}.
	\newblock \DOIprefix\doi{10.1063/1.556010}.
	%Type = Article
	\bibitem[{{Atkinson} and {Welge}(1972)}]{Atkinson72}
	\bibinfo{author}{{Atkinson}, R.}, \bibinfo{author}{{Welge}, K.H.},
	\bibinfo{year}{1972}.
	\newblock \bibinfo{title}{{Temperature Dependence of O($^{1}$S) 
	Deactivation by
			CO$_{2}$, O$_{2}$, N$_{2}$, and Ar}}.
	\newblock \bibinfo{journal}{\jcp} \bibinfo{volume}{57},
	\bibinfo{pages}{3689--3693}.
	\newblock \DOIprefix\doi{10.1063/1.1678829}.
	%Type = Article
	\bibitem[{{Benna} and {Elrod}(2018)}]{Benna18}
	\bibinfo{author}{{Benna}, M.}, \bibinfo{author}{{Elrod}, M.},
	\bibinfo{year}{2018}.
	\newblock \bibinfo{title}{Mars atmosphere and volatile evolution ({MAVEN})
		mission, neutral gas and ion mass spectrometer ({NGIMS})}.
	\newblock \bibinfo{journal}{NGIMS PDS Software Interface Specification,
		Revision 1.9} .
	%Type = Article
	\bibitem[{Berrington and Burke(1981)}]{Berrington81}
	\bibinfo{author}{Berrington, K.A.}, \bibinfo{author}{Burke, P.G.},
	\bibinfo{year}{1981}.
	\newblock \bibinfo{title}{{Effective collision strengths for forbidden
			transitions in e-N and e-O scattering}}.
	\newblock \bibinfo{journal}{Planetary and Space Science} 
	\bibinfo{volume}{29},
	\bibinfo{pages}{377 -- 381}.
	\newblock \DOIprefix\doi{10.1016/0032-0633(81)90026-X}.
	%Type = Article
	\bibitem[{{Bhardwaj} and {Raghuram}(2012)}]{Bhardwaj12}
	\bibinfo{author}{{Bhardwaj}, A.}, \bibinfo{author}{{Raghuram}, S.},
	\bibinfo{year}{2012}.
	\newblock \bibinfo{title}{{A coupled chemistry-emission model for atomic 
	oxygen
			green and red-doublet emissions in the comet C/1996 B2 Hyakutake}}.
	\newblock \bibinfo{journal}{\apj} \bibinfo{volume}{748}, 
	\bibinfo{pages}{13}.
	\newblock \DOIprefix\doi{10.1088/0004-637X/748/1/13}.
	%Type = Misc
	\bibitem[{{Burkholder} et~al.(2015){Burkholder}, {Sander}, {Abbatt}, 
	{Barker},
		{Huie}, {Kolb}, {Kurylo}, {Orkin}, {Wilmouth} and {Wine}}]{Burkholder15}
	\bibinfo{author}{{Burkholder}, J.B.}, \bibinfo{author}{{Sander}, S.P.},
	\bibinfo{author}{{Abbatt}, J.}, \bibinfo{author}{{Barker}, J.R.},
	\bibinfo{author}{{Huie}, R.E.}, \bibinfo{author}{{Kolb}, C.E.},
	\bibinfo{author}{{Kurylo}, M.J.}, \bibinfo{author}{{Orkin}, V.L.},
	\bibinfo{author}{{Wilmouth}, D.M.}, \bibinfo{author}{{Wine}, P.H.},
	\bibinfo{year}{2015}.
	\newblock \bibinfo{title}{Chemical kinetics and photochemical data for use 
	in
		atmospheric studies, evaluation no. 18}.
	%Type = Article
	\bibitem[{Capetanakis et~al.(1993)Capetanakis, Sondermann, H\"oser and
		Stuhl}]{Capetanakis93}
	\bibinfo{author}{Capetanakis, F.P.}, \bibinfo{author}{Sondermann, F.},
	\bibinfo{author}{H\"oser, S.}, \bibinfo{author}{Stuhl, F.},
	\bibinfo{year}{1993}.
	\newblock \bibinfo{title}{{Temperature dependence of the quenching of 
	O($^1$S)
			by simple inorganic molecules}}.
	\newblock \bibinfo{journal}{\jcp} \bibinfo{volume}{98}, 
	\bibinfo{pages}{7883}.
	\newblock \DOIprefix\doi{10.1063/1.464596}.
	%Type = Article
	\bibitem[{{Eparvier} et~al.(2015){Eparvier}, {Chamberlin}, {Woods} and
		{Thiemann}}]{Eparvier15}
	\bibinfo{author}{{Eparvier}, F.G.}, \bibinfo{author}{{Chamberlin}, P.C.},
	\bibinfo{author}{{Woods}, T.N.}, \bibinfo{author}{{Thiemann}, E.M.B.},
	\bibinfo{year}{2015}.
	\newblock \bibinfo{title}{{The Solar Extreme Ultraviolet Monitor for 
	MAVEN}}.
	\newblock \bibinfo{journal}{\ssr} \bibinfo{volume}{195},
	\bibinfo{pages}{293--301}.
	\newblock \DOIprefix\doi{10.1007/s11214-015-0195-2}.
	%Type = Article
	\bibitem[{{Fox}(2004)}]{Fox04}
	\bibinfo{author}{{Fox}, J.L.}, \bibinfo{year}{2004}.
	\newblock \bibinfo{title}{{Response of the Martian thermosphere/ionosphere 
	to
			enhanced fluxes of solar soft X rays}}.
	\newblock \bibinfo{journal}{J. Geophys. Res. (Space Physics)}
	\bibinfo{volume}{109}, \bibinfo{pages}{A11310}.
	\newblock \DOIprefix\doi{10.1029/2004JA010380}.
	%Type = Article
	\bibitem[{Fox and Dalgarno(1979)}]{Fox79}
	\bibinfo{author}{Fox, J.L.}, \bibinfo{author}{Dalgarno, A.},
	\bibinfo{year}{1979}.
	\newblock \bibinfo{title}{{Ionization, luminosity, and heating of the upper
			atmosphere of Mars}}.
	\newblock \bibinfo{journal}{\jgr} \bibinfo{volume}{84}, 
	\bibinfo{pages}{7315 --
		7333}.
	\newblock \DOIprefix\doi{10.1029/JA084iA12p07315}.
	%Type = Article
	\bibitem[{{Fox} and {Ha{\'c}}(2009)}]{Fox09}
	\bibinfo{author}{{Fox}, J.L.}, \bibinfo{author}{{Ha{\'c}}, A.B.},
	\bibinfo{year}{2009}.
	\newblock \bibinfo{title}{{Photochemical escape of oxygen from Mars: A
			comparison of the exobase approximation to a Monte Carlo method}}.
	\newblock \bibinfo{journal}{\icarus} \bibinfo{volume}{204},
	\bibinfo{pages}{527--544}.
	\newblock \DOIprefix\doi{10.1016/j.icarus.2009.07.005}.
	%Type = Article
	\bibitem[{{Froese Fischer} and {Tachiev}(2004)}]{Froese04}
	\bibinfo{author}{{Froese Fischer}, C.}, \bibinfo{author}{{Tachiev}, G.},
	\bibinfo{year}{2004}.
	\newblock \bibinfo{title}{{Breit-Pauli energy levels, lifetimes, and 
	transition
			probabilities for the beryllium-like to neon-like sequences}}.
	\newblock \bibinfo{journal}{Atomic Data and Nuclear Data Tables}
	\bibinfo{volume}{87}, \bibinfo{pages}{1--184}.
	\newblock \DOIprefix\doi{10.1016/j.adt.2004.02.001}.
	%Type = Article
	\bibitem[{Gao and Ng(2019)}]{Gao19}
	\bibinfo{author}{Gao, H.}, \bibinfo{author}{Ng, C.Y.}, \bibinfo{year}{2019}.
	\newblock \bibinfo{title}{Quantum state-to-state vacuum ultraviolet
		photodissociation dynamics of small molecules}.
	\newblock \bibinfo{journal}{Chinese Journal of Chemical Physics}
	\bibinfo{volume}{32}, \bibinfo{pages}{23--34}.
	\newblock \DOIprefix\doi{10.1063/1674-0068/cjcp1812290}.
	%Type = Article
	\bibitem[{{G{\'e}rard} et~al.(2020){G{\'e}rard}, Aoki, Willame, Gkouvelis,
		Depiesse, Thomas, Ristic, Vandaele, Daerden, Hubert, Mason, Patel,
		{L{\'o}pez-Moreno}, Bellucci, {L{\'o}pez-Valverde} and 
		Beeckman}]{Gerard20}
	\bibinfo{author}{{G{\'e}rard}, J.C.}, \bibinfo{author}{Aoki, S.},
	\bibinfo{author}{Willame, Y.}, \bibinfo{author}{Gkouvelis, L.},
	\bibinfo{author}{Depiesse, C.}, \bibinfo{author}{Thomas, I.R.},
	\bibinfo{author}{Ristic, B.}, \bibinfo{author}{Vandaele, A.C.},
	\bibinfo{author}{Daerden, F.}, \bibinfo{author}{Hubert, B.},
	\bibinfo{author}{Mason, J.}, \bibinfo{author}{Patel, M.R.},
	\bibinfo{author}{{L{\'o}pez-Moreno}, J.J.}, \bibinfo{author}{Bellucci, G.},
	\bibinfo{author}{{L{\'o}pez-Valverde}, M.A.}, \bibinfo{author}{Beeckman, 
	B.},
	\bibinfo{year}{2020}.
	\newblock \bibinfo{title}{{Detection of green line emission in the dayside
			atmosphere of Mars from NOMAD-TGO observations}}.
	\newblock \bibinfo{journal}{\nat} \bibinfo{volume}{124}, 
	\bibinfo{pages}{5--8}.
	\newblock \DOIprefix\doi{10.1029/2019JA026596}.
	%Type = Article
	\bibitem[{{Gkouvelis} et~al.(2018){Gkouvelis}, {G{\'e}rard}, {Ritter},
		{Hubert}, {Schneider} and {Jain}}]{Gkouvelis18}
	\bibinfo{author}{{Gkouvelis}, L.}, \bibinfo{author}{{G{\'e}rard}, J.C.},
	\bibinfo{author}{{Ritter}, B.}, \bibinfo{author}{{Hubert}, B.},
	\bibinfo{author}{{Schneider}, N.M.}, \bibinfo{author}{{Jain}, S.K.},
	\bibinfo{year}{2018}.
	\newblock \bibinfo{title}{{The O($^{1}$S) 297.2 nm Dayglow Emission: A 
	Tracer
			of CO$_{2}$ Density Variations in the Martian Lower Thermosphere}}.
	\newblock \bibinfo{journal}{Journal of Geophysical Research (Planets)}
	\bibinfo{volume}{123}, \bibinfo{pages}{3119--3132}.
	\newblock \DOIprefix\doi{10.1029/2018JE005709}.
	%Type = Article
	\bibitem[{{Gkouvelis} et~al.(2020){Gkouvelis}, {G{\'e}rard}, {Ritter},
		{Hubert}, {Schneider} and {Jain}}]{Gkouvelis20}
	\bibinfo{author}{{Gkouvelis}, L.}, \bibinfo{author}{{G{\'e}rard}, J.C.},
	\bibinfo{author}{{Ritter}, B.}, \bibinfo{author}{{Hubert}, B.},
	\bibinfo{author}{{Schneider}, N.M.}, \bibinfo{author}{{Jain}, S.K.},
	\bibinfo{year}{2020}.
	\newblock \bibinfo{title}{{Airglow remote sensing of the seasonal variation 
	of
			the Martian upper atmosphere: MAVEN limb observations and model 
			comparison}}.
	\newblock \bibinfo{journal}{\icarus} \bibinfo{volume}{341},
	\bibinfo{pages}{113666}.
	\newblock \DOIprefix\doi{10.1016/j.icarus.2020.113666}.
	%Type = Article
	\bibitem[{{Gronoff} et~al.(2012a){Gronoff}, {Wedlund}, {Mertens},
		{Barth{\'e}lemy}, {Lillis} and {Witasse}}]{Gronoff12b}
	\bibinfo{author}{{Gronoff}, G.}, \bibinfo{author}{{Wedlund}, C.S.},
	\bibinfo{author}{{Mertens}, C.J.}, \bibinfo{author}{{Barth{\'e}lemy}, M.},
	\bibinfo{author}{{Lillis}, R.J.}, \bibinfo{author}{{Witasse}, O.},
	\bibinfo{year}{2012}a.
	\newblock \bibinfo{title}{{Computing uncertainties in ionosphere-airglow
			models: II. The Martian airglow}}.
	\newblock \bibinfo{journal}{Journal of Geophysical Research (Space Physics)}
	\bibinfo{volume}{117}, \bibinfo{pages}{A05309}.
	\newblock \DOIprefix\doi{10.1029/2011JA017308}.
	%Type = Article
	\bibitem[{{Gronoff} et~al.(2012b){Gronoff}, {Wedlund}, {Mertens} and
		{Lillis}}]{Gronoff12a}
	\bibinfo{author}{{Gronoff}, G.}, \bibinfo{author}{{Wedlund}, C.S.},
	\bibinfo{author}{{Mertens}, C.J.}, \bibinfo{author}{{Lillis}, R.J.},
	\bibinfo{year}{2012}b.
	\newblock \bibinfo{title}{{Computing uncertainties in ionosphere-airglow
			models: I. Electron flux and species production uncertainties for 
			Mars}}.
	\newblock \bibinfo{journal}{Journal of Geophysical Research (Space Physics)}
	\bibinfo{volume}{117}, \bibinfo{pages}{A04306}.
	\newblock \DOIprefix\doi{10.1029/2011JA016930}.
	%Type = Article
	\bibitem[{{Guberman}(1988)}]{Guberman88}
	\bibinfo{author}{{Guberman}, S.L.}, \bibinfo{year}{1988}.
	\newblock \bibinfo{title}{{The production of O($^1$D) from dissociative
			recombination of O$_2^+$}}.
	\newblock \bibinfo{journal}{\planss} ,
	\bibinfo{pages}{47--53}\DOIprefix\doi{10.1016/0032-0633(88)90145-6}.
	%Type = Article
	\bibitem[{Huestis and Slanger(2006)}]{Huestis06}
	\bibinfo{author}{Huestis, D.L.}, \bibinfo{author}{Slanger, T.G.},
	\bibinfo{year}{2006}.
	\newblock \bibinfo{title}{{DPS}}.
	\newblock \bibinfo{journal}{American Astronomical Society}
	\bibinfo{volume}{38}, \bibinfo{pages}{62.20}.
	%Type = Article
	\bibitem[{{Huestis} et~al.(2010){Huestis}, {Slanger}, {Sharpee} and
		{Fox}}]{Huestis10}
	\bibinfo{author}{{Huestis}, D.L.}, \bibinfo{author}{{Slanger}, T.G.},
	\bibinfo{author}{{Sharpee}, B.D.}, \bibinfo{author}{{Fox}, J.L.},
	\bibinfo{year}{2010}.
	\newblock \bibinfo{title}{{Chemical origins of the Mars ultraviolet 
	dayglow}}.
	\newblock \bibinfo{journal}{Faraday Discuss.} \bibinfo{volume}{147},
	\bibinfo{pages}{307}.
	\newblock \DOIprefix\doi{10.1039/c003456h}.
	%Type = Phdthesis
	\bibitem[{Jain(2013)}]{Jain13b}
	\bibinfo{author}{Jain, S.K.}, \bibinfo{year}{2013}.
	\newblock \bibinfo{title}{{Dayglow emissions on Mars and Venus, Phd. 
	Thesis}}.
	\newblock Ph.D. thesis. Cochin University of Science and Technology, India.
	\bibinfo{address}{India}.
	\newblock \URLprefix \url{https://dyuthi. 
	cusat.ac.in/jspui/handle/purl/3688}.
	%Type = Article
	\bibitem[{{Jain} and {Bhardwaj}(2012)}]{Jain12}
	\bibinfo{author}{{Jain}, S.K.}, \bibinfo{author}{{Bhardwaj}, A.},
	\bibinfo{year}{2012}.
	\newblock \bibinfo{title}{{Impact of solar EUV flux on CO Cameron band and
			CO$_2$ UV doublet emissions in the dayglow of Mars}}.
	\newblock \bibinfo{journal}{\planss} \bibinfo{volume}{64}, 
	\bibinfo{pages}{110
		-- 122}.
	%Type = Article
	\bibitem[{{Jain} et~al.(2015){Jain}, {Stewart}, {Schneider}, {Deighan},
		{Stiepen}, {Evans}, {Stevens}, {Chaffin}, {Crismani}, {McClintock}, 
		{Clarke},
		{Holsclaw}, {Lo}, {Lef{\`e}vre}, {Montmessin}, {Thiemann}, {Eparvier} 
		and
		{Jakosky}}]{Jain15}
	\bibinfo{author}{{Jain}, S.K.}, \bibinfo{author}{{Stewart}, A.I.F.},
	\bibinfo{author}{{Schneider}, N.M.}, \bibinfo{author}{{Deighan}, J.},
	\bibinfo{author}{{Stiepen}, A.}, \bibinfo{author}{{Evans}, J.S.},
	\bibinfo{author}{{Stevens}, M.H.}, \bibinfo{author}{{Chaffin}, M.S.},
	\bibinfo{author}{{Crismani}, M.}, \bibinfo{author}{{McClintock}, W.E.},
	\bibinfo{author}{{Clarke}, J.T.}, \bibinfo{author}{{Holsclaw}, G.M.},
	\bibinfo{author}{{Lo}, D.Y.}, \bibinfo{author}{{Lef{\`e}vre}, F.},
	\bibinfo{author}{{Montmessin}, F.}, \bibinfo{author}{{Thiemann}, E.M.B.},
	\bibinfo{author}{{Eparvier}, F.}, \bibinfo{author}{{Jakosky}, B.M.},
	\bibinfo{year}{2015}.
	\newblock \bibinfo{title}{{The structure and variability of Mars upper
			atmosphere as seen in MAVEN/IUVS dayglow observations}}.
	\newblock \bibinfo{journal}{\grl} \bibinfo{volume}{42},
	\bibinfo{pages}{9023--9030}.
	\newblock \DOIprefix\doi{10.1002/2015GL065419}.
	%Type = Article
	\bibitem[{{Krauss} and {Neumann}(1975)}]{Krauss75}
	\bibinfo{author}{{Krauss}, M.}, \bibinfo{author}{{Neumann}, D.},
	\bibinfo{year}{1975}.
	\newblock \bibinfo{title}{{On the interaction of O($^{1}$S) with 
	O($^{3}$P)}}.
	\newblock \bibinfo{journal}{Chemical Physics Letters} \bibinfo{volume}{36},
	\bibinfo{pages}{372--374}.
	\newblock \DOIprefix\doi{10.1016/0009-2614(75)80259-4}.
	%Type = Article
	\bibitem[{{Leblanc} et~al.(2006){Leblanc}, {Chaufray}, {Lilensten}, 
	{Witasse}
		and {Bertaux}}]{Leblanc06}
	\bibinfo{author}{{Leblanc}, F.}, \bibinfo{author}{{Chaufray}, J.Y.},
	\bibinfo{author}{{Lilensten}, J.}, \bibinfo{author}{{Witasse}, O.},
	\bibinfo{author}{{Bertaux}, J.L.}, \bibinfo{year}{2006}.
	\newblock \bibinfo{title}{{Martian dayglow as seen by the SPICAM UV
			spectrograph on Mars Express}}.
	\newblock \bibinfo{journal}{Journal of Geophysical Research (Planets)}
	\bibinfo{volume}{111}, \bibinfo{pages}{E09S11}.
	\newblock \DOIprefix\doi{10.1029/2005JE002664}.
	%Type = Article
	\bibitem[{{Lu} et~al.(2015){Lu}, {Chang}, {Benitez}, {Luo}, {Houria}, 
	{Ayari},
		{Al Mogren}, {Hochlaf}, {Jackson} and {Ng}}]{Lu15}
	\bibinfo{author}{{Lu}, Z.}, \bibinfo{author}{{Chang}, Y.C.},
	\bibinfo{author}{{Benitez}, Y.}, \bibinfo{author}{{Luo}, Z.},
	\bibinfo{author}{{Houria}, A.B.}, \bibinfo{author}{{Ayari}, T.},
	\bibinfo{author}{{Al Mogren}, M.M.}, \bibinfo{author}{{Hochlaf}, M.},
	\bibinfo{author}{{Jackson}, W.M.}, \bibinfo{author}{{Ng}, C.Y.},
	\bibinfo{year}{2015}.
	\newblock \bibinfo{title}{{State-to-state vacuum ultraviolet 
	photodissociation
			study of CO$_2$ on the formation of state-correlated 
			CO(X$^1\Sigma^+$; v)
			with O($^1$D) and O($^1$S) photoproducts at 11.95-12.22 eV}}.
	\newblock \bibinfo{journal}{Physical Chemistry Chemical Physics 
	(Incorporating
		Faraday Transactions)} \bibinfo{volume}{17}, 
		\bibinfo{pages}{11752--11762}.
	\newblock \DOIprefix\doi{10.1039/C5CP01321F}.
	%Type = Article
	\bibitem[{{Nu{\~n}ez-Reyes} and {Hickson}(2018)}]{Reyes18}
	\bibinfo{author}{{Nu{\~n}ez-Reyes}, D.}, \bibinfo{author}{{Hickson}, K.M.},
	\bibinfo{year}{2018}.
	\newblock \bibinfo{title}{{Kinetics of the Gas-Phase O($^1$D) + CO$_2$ and
			C($^1$D) + CO$_2$ Reactions over the 50--296 K range}}.
	\newblock \bibinfo{journal}{Journal of Physical Chemistry A}
	\bibinfo{volume}{122}, \bibinfo{pages}{4002--4008}.
	\newblock \DOIprefix\doi{10.1021/acs.jpca.8b01964}.
	%Type = Article
	\bibitem[{{Raghuram} and {Bhardwaj}(2012)}]{Raghuram12}
	\bibinfo{author}{{Raghuram}, S.}, \bibinfo{author}{{Bhardwaj}, A.},
	\bibinfo{year}{2012}.
	\newblock \bibinfo{title}{{Model for the production of CO Cameron band 
	emission
			in Comet 1P/Halley}}.
	\newblock \bibinfo{journal}{\planss} \bibinfo{volume}{63},
	\bibinfo{pages}{139--149}.
	\newblock \DOIprefix\doi{10.1016/j.pss.2011.11.011},
	\href{http://arxiv.org/abs/1201.6291}{\tt arXiv:1201.6291}.
	%Type = Article
	\bibitem[{{Raghuram} and {Bhardwaj}(2020)}]{Raghuram20a}
	\bibinfo{author}{{Raghuram}, S.}, \bibinfo{author}{{Bhardwaj}, A.},
	\bibinfo{year}{2020}.
	\newblock \bibinfo{title}{{CO+ first-negative band emission: A tracer for 
	CO in
			the Martian upper atmosphere}}.
	\newblock \bibinfo{journal}{\aap} \bibinfo{volume}{639}, 
	\bibinfo{pages}{A60}.
	\newblock \DOIprefix\doi{10.1051/0004-6361/202038147}.
	%Type = Article
	\bibitem[{{Raghuram} et~al.(2020){Raghuram}, {Hutsem{\'e}kers}, {Opitom},
		{Jehin}, {Bhardwaj} and {Manfroid}}]{Raghuram20b}
	\bibinfo{author}{{Raghuram}, S.}, \bibinfo{author}{{Hutsem{\'e}kers}, D.},
	\bibinfo{author}{{Opitom}, C.}, \bibinfo{author}{{Jehin}, E.},
	\bibinfo{author}{{Bhardwaj}, A.}, \bibinfo{author}{{Manfroid}, J.},
	\bibinfo{year}{2020}.
	\newblock \bibinfo{title}{{Forbidden atomic carbon, nitrogen, and oxygen
			emission lines in the water-poor comet C/2016 R2 (Pan-STARRS)}}.
	\newblock \bibinfo{journal}{\aap} \bibinfo{volume}{635}, 
	\bibinfo{pages}{A108}.
	\newblock \DOIprefix\doi{10.1051/0004-6361/201936713},
	\href{http://arxiv.org/abs/2001.03315}{\tt arXiv:2001.03315}.
	%Type = Article
	\bibitem[{{Ros{\'e}n} et~al.(1998){Ros{\'e}n}, {Peverall}, {Larsson}, {Le
			Padellec}, {Semaniak}, {Larson}, {Str{\"o}mholm}, {van der Zande}, 
			{Danared}
		and {Dunn}}]{Rosen98}
	\bibinfo{author}{{Ros{\'e}n}, S.}, \bibinfo{author}{{Peverall}, R.},
	\bibinfo{author}{{Larsson}, M.}, \bibinfo{author}{{Le Padellec}, A.},
	\bibinfo{author}{{Semaniak}, J.}, \bibinfo{author}{{Larson}, {\AA}.},
	\bibinfo{author}{{Str{\"o}mholm}, C.}, \bibinfo{author}{{van der Zande},
		W.J.}, \bibinfo{author}{{Danared}, H.}, \bibinfo{author}{{Dunn}, G.H.},
	\bibinfo{year}{1998}.
	\newblock \bibinfo{title}{{Absolute cross sections and final-state
			distributions for dissociative recombination and excitation of 
			CO$^{+}$(v=0)
			using an ion storage ring}}.
	\newblock \bibinfo{journal}{\pra} \bibinfo{volume}{57},
	\bibinfo{pages}{4462--4471}.
	\newblock \DOIprefix\doi{10.1103/PhysRevA.57.4462}.
	%Type = Article
	\bibitem[{Schofield(1978)}]{Schofield78}
	\bibinfo{author}{Schofield, K.}, \bibinfo{year}{1978}.
	\newblock \bibinfo{title}{{Rate constants for the gaseous interaction of
			O(2$^1$D$_2$) and O(2$^1$S$_0$) - a critical evaluation}}.
	\newblock \bibinfo{journal}{Journal of Photochemistry} \bibinfo{volume}{9},
	\bibinfo{pages}{55 -- 68}.
	\newblock \DOIprefix\doi{10.1016/0047-2670(78)87006-3}.
	%Type = Article
	\bibitem[{{Simon} et~al.(2009){Simon}, {Witasse}, {Leblanc}, {Gronoff} and
		{Bertaux}}]{Simon09}
	\bibinfo{author}{{Simon}, C.}, \bibinfo{author}{{Witasse}, O.},
	\bibinfo{author}{{Leblanc}, F.}, \bibinfo{author}{{Gronoff}, G.},
	\bibinfo{author}{{Bertaux}, J.L.}, \bibinfo{year}{2009}.
	\newblock \bibinfo{title}{{Dayglow on Mars: Kinetic modelling with SPICAM 
	UV
			limb data}}.
	\newblock \bibinfo{journal}{\planss} \bibinfo{volume}{57},
	\bibinfo{pages}{1008--1021}.
	\newblock \DOIprefix\doi{10.1016/j.pss.2008.08.012}.
	%Type = Article
	\bibitem[{{Song} et~al.(2014){Song}, {Gao}, {Chang}, {Lu}, {Ng} and
		{Jackson}}]{Song14}
	\bibinfo{author}{{Song}, Y.}, \bibinfo{author}{{Gao}, H.},
	\bibinfo{author}{{Chang}, Y.C.}, \bibinfo{author}{{Lu}, Z.},
	\bibinfo{author}{{Ng}, C.Y.}, \bibinfo{author}{{Jackson}, W.M.},
	\bibinfo{year}{2014}.
	\newblock \bibinfo{title}{{Photodissociation of CO$_2$ between 13.540 eV and
			13.678 eV}}.
	\newblock \bibinfo{journal}{Physical Chemistry Chemical Physics 
	(Incorporating
		Faraday Transactions)} \bibinfo{volume}{16}, \bibinfo{pages}{563}.
	\newblock \DOIprefix\doi{10.1039/C3CP53250J}.
	%Type = Article
	\bibitem[{{Stewart}(1972)}]{Stewart72a}
	\bibinfo{author}{{Stewart}, A.I.}, \bibinfo{year}{1972}.
	\newblock \bibinfo{title}{{Mariner 6 and 7 Ultraviolet Spectrometer 
	Experiment:
			Implications of CO$_{2}^{+}$, CO and O Airglow}}.
	\newblock \bibinfo{journal}{\jgr} \bibinfo{volume}{77}, \bibinfo{pages}{54}.
	\newblock \DOIprefix\doi{10.1029/JA077i001p00054}.
	%Type = Article
	\bibitem[{{Streit} et~al.(1976){Streit}, {Howard}, {Schmeltekopf}, 
	{Davidson}
		and {Schiff}}]{Streit76}
	\bibinfo{author}{{Streit}, G.E.}, \bibinfo{author}{{Howard}, C.J.},
	\bibinfo{author}{{Schmeltekopf}, A.L.}, \bibinfo{author}{{Davidson}, J.A.},
	\bibinfo{author}{{Schiff}, H.I.}, \bibinfo{year}{1976}.
	\newblock \bibinfo{title}{{Temperature dependence of O($^1$D) rate constants
			for reactions with O$_2$, N$_2$, CO$_2$, O$_3$, and H$_2$O}}.
	\newblock \bibinfo{journal}{\jcp} \bibinfo{volume}{65},
	\bibinfo{pages}{4761--4764}.
	\newblock \DOIprefix\doi{10.1063/1.432930}.
	%Type = Article
	\bibitem[{Sutradhar et~al.(2017)Sutradhar, Samanta, Samanta and
		Reisler}]{Sutradhar17}
	\bibinfo{author}{Sutradhar, S.}, \bibinfo{author}{Samanta, B.R.},
	\bibinfo{author}{Samanta, A.K.}, \bibinfo{author}{Reisler, H.},
	\bibinfo{year}{2017}.
	\newblock \bibinfo{title}{{Temperature dependence of the photodissociation 
	of
			CO$_2$ from high vibrational levels: 205-230 nm imaging studies of
			CO(X$^1\Sigma^+$) and O($^3$P, $^1$D) products}}.
	\newblock \bibinfo{journal}{\jcp} \bibinfo{volume}{147},
	\bibinfo{pages}{013916}.
	\newblock \DOIprefix\doi{10.1063/1.4979952}.
	%Type = Article
	\bibitem[{{Thiemann} et~al.(2017){Thiemann}, {Chamberlin}, {Eparvier},
		{Templeman}, {Woods}, {Bougher} and {Jakosky}}]{Thiemann17}
	\bibinfo{author}{{Thiemann}, E.M.B.}, \bibinfo{author}{{Chamberlin}, P.C.},
	\bibinfo{author}{{Eparvier}, F.G.}, \bibinfo{author}{{Templeman}, B.},
	\bibinfo{author}{{Woods}, T.N.}, \bibinfo{author}{{Bougher}, S.W.},
	\bibinfo{author}{{Jakosky}, B.M.}, \bibinfo{year}{2017}.
	\newblock \bibinfo{title}{{The MAVEN EUVM model of solar spectral irradiance
			variability at Mars: Algorithms and results}}.
	\newblock \bibinfo{journal}{Journal of Geophysical Research (Space Physics)}
	\bibinfo{volume}{122}, \bibinfo{pages}{2748--2767}.
	\newblock \DOIprefix\doi{10.1002/2016JA023512}.
	%Type = Article
	\bibitem[{{Viggiano} et~al.(2005){Viggiano}, {Ehlerding}, {Hellberg}, 
	{Thomas},
		{Zhaunerchyk}, {Geppert}, {Montaigne}, {Larsson}, {Kaminska} and
		{{\"O}sterdahl}}]{Viggiano05}
	\bibinfo{author}{{Viggiano}, A.A.}, \bibinfo{author}{{Ehlerding}, A.},
	\bibinfo{author}{{Hellberg}, F.}, \bibinfo{author}{{Thomas}, R.D.},
	\bibinfo{author}{{Zhaunerchyk}, V.}, \bibinfo{author}{{Geppert}, W.D.},
	\bibinfo{author}{{Montaigne}, H.}, \bibinfo{author}{{Larsson}, M.},
	\bibinfo{author}{{Kaminska}, M.}, \bibinfo{author}{{{\"O}sterdahl}, F.},
	\bibinfo{year}{2005}.
	\newblock \bibinfo{title}{{Rate constants and branching ratios for the
			dissociative recombination of CO$_{2}$$^{+}$}}.
	\newblock \bibinfo{journal}{\jcp} \bibinfo{volume}{122},
	\bibinfo{pages}{226101--226101}.
	\newblock \DOIprefix\doi{10.1063/1.1926283}.
	%Type = Book
	\bibitem[{Wiese et~al.(1996)Wiese, Fuhr and Deters}]{Wiese96}
	\bibinfo{author}{Wiese, W.L.}, \bibinfo{author}{Fuhr, J.R.},
	\bibinfo{author}{Deters, T.M.}, \bibinfo{year}{1996}.
	\newblock \bibinfo{title}{{Atomic transition probabilities of carbon, 
	nitrogen,
			and oxygen: A critical data compilation}}.
	\newblock \bibinfo{publisher}{Am. Chem. Soc., Washington, D. C.}
	%Type = Article
	\bibitem[{{Yee} et~al.(1990){Yee}, {Guberman} and {Dalgarno}}]{Yee90}
	\bibinfo{author}{{Yee}, J.H.}, \bibinfo{author}{{Guberman}, S.L.},
	\bibinfo{author}{{Dalgarno}, A.}, \bibinfo{year}{1990}.
	\newblock \bibinfo{title}{{Collisional quenching of O($^{1}$D) by 
	O($^{3}$P)}}.
	\newblock \bibinfo{journal}{\planss} \bibinfo{volume}{38},
	\bibinfo{pages}{647--652}.
	\newblock \DOIprefix\doi{10.1016/0032-0633(90)90071-W}.
	
\end{thebibliography}


%Author biography without author photo.
%Author biography. Author biography. Author biography.
%Author biography. Author biography. Author biography.
%Author biography. Author biography. Author biography.
%Author biography. Author biography. Author biography.
%Author biography. Author biography. Author biography.
%Author biography. Author biography. Author biography.
%Author biography. Author biography. Author biography.
%Author biography. Author biography. Author biography.
%Author biography. Author biography. Author biography.
%\bio{figs/pic1}
%Author biography with author photo.
%Author biography. Author biography. Author biography.
%Author biography. Author biography. Author biography.
%Author biography. Author biography. Author biography.
%Author biography. Author biography. Author biography.
%Author biography. Author biography. Author biography.
%Author biography. Author biography. Author biography.
%Author biography. Author biography. Author biography.
%Author biography. Author biography. Author biography.
%Author biography. Author biography. Author biography.
%\bio{figs/pic1}
%Author biography with author photo.
%Author biography. Author biography. Author biography.
%Author biography. Author biography. Author biography.
%Author biography. Author biography. Author biography.
%Author biography. Author biography. Author biography.

\end{document}